\documentclass[useAMS,usenatbib]{mn2e}
\usepackage{graphicx}
\usepackage{graphics}
\usepackage{rotating}
\usepackage{amsmath}
\usepackage{amssymb}

\usepackage{times}

\newcommand{\changes}[1]{{#1}}

\title[AT20G Source Catalogue]
      {The Australia Telescope 20 GHz Survey: The Source Catalogue}
\author[Murphy et al.]{
\parbox[t]{\textwidth}
{Tara Murphy$^{1,2}$\thanks{E-mail: tara@physics.usyd.edu.au}, 
Elaine M.\ Sadler$^1$, Ronald D.\ Ekers$^3$,
Marcella Massardi$^4$, Paul J.\ Hancock$^1$, 
Elizabeth Mahony$^{1,3}$, Roberto Ricci$^5$, 
Sarah Burke-Spolaor$^{3,6}$, Mark Calabretta$^3$, 
Rajan Chhetri$^{3,7}$, Gianfranco De Zotti$^{4,8}$, 
Philip G.\ Edwards$^3$, Jennifer A. Ekers$^3$,  Carole A.\ Jackson$^3$, 
Michael J.\ Kesteven$^3$, Emma Lindley$^{1}$,  Katherine Newton-McGee$^{1,3}$, 
Chris Phillips$^3$, Paul Roberts$^3$, Robert J.\ Sault$^{9}$, 
Lister Staveley-Smith$^{10}$, Ravi Subrahmanyan$^{11}$, Mark A.\ Walker$^{12}$, 
and Warwick E.\ Wilson$^3$}
\vspace*{6pt} \\
$^{1}$Sydney Institute for Astronomy, School of Physics, The University of Sydney, NSW 2006, Australia\\
$^{2}$School of Information Technologies, The University of Sydney, NSW 2006, Australia\\
$^{3}$Australia Telescope National Facility, CSIRO, P.O. Box 76,
Epping, NSW 1710, Australia \\
$^{4}$INAF, Osservatorio Astronomico di Padova, Vicolo
dell'Osservatorio 5, I-35122 Padova, Italy \\
$^{5}$Department of Physics and Astronomy, University of Calgary,
2500 University Drive NW Calgary, AB, Canada\\
$^{6}$Swinburne University of Technology, P.O.\ Box 218, Hawthorn, Vic 3122, Australia\\
$^{7}$School of Physics, The University of New South Wales, NSW 2052, Australia\\
$^{8}$SISSA/ISAS, Via Beirut 2--4, I-34014 Trieste, Italy\\
$^{9}$School of Physics, The University of Melbourne, Victoria 3010, Australia\\
$^{10}$School of Physics, The University of Western Australia, 35 Stirling Highway Crawley, WA 6009, Australia\\
$^{11}$Raman Research Institute, Sadashivanagar, Bangalore 560080, India\\
$^{12}$Manly Astrophysics, 3/22 Cliff St., Manly 2095, Australia
}

\date{Accepted 0000 December 08. Received 0000 December 08; in original form 0000 December 08}

\pagerange{\pageref{firstpage}--\pageref{lastpage}} 
\pubyear{2008}

\begin{document}

\maketitle

\label{firstpage}

\begin{abstract}
We present the full source catalogue from the Australia Telescope 20~GHz (AT20G) Survey.  
The AT20G is a blind radio survey carried out at 20~GHz with the Australia 
Telescope Compact Array (ATCA) from 2004 to 2008, and covers the whole sky south of 
declination 0\degr. The AT20G source catalogue presented here is an order of magnitude 
larger than any previous catalogue of high-frequency radio sources, and includes 5890 
sources above a 20~GHz flux-density limit of 40\,mJy.  All AT20G sources have total intensity 
and polarisation measured at 20~GHz, and most sources south of declination $-15\degr$ 
also have near-simultaneous flux-density measurements at 5 and 8~GHz. 
A total of 1559 sources were detected in polarised total intensity at one or more 
of the three frequencies.

The completeness of the AT20G source catalogue is 91 per cent above 100~mJy\,beam$^{-1}$ and 79 per cent above
50~mJy\,beam$^{-1}$ in regions south of declination $-15\degr$. North of $-15\degr$, some observations 
of sources between $14-20$~hr in right ascension were lost due to bad weather and could not be repeated, 
so the catalogue completeness is lower in this region. 
Each detected source was visually inspected as part of our quality control process, and so the 
reliability of the final catalogue is essentially 100 per cent.

We detect a small but significant population of non-thermal sources that are either undetected or have 
only weak detections in low-frequency catalogues. We introduce the term Ultra-Inverted Spectrum 
(UIS) to describe these radio sources, which have a spectral index $\alpha(5,20)>+0.7$ and which constitute roughly 
1.2 per cent of the AT20G sample.

The 20~GHz flux densities measured for the strongest AT20G sources are in excellent 
agreement with the WMAP 5-year source catalogue of \citet{wright09},
and we find that the WMAP source catalogue is close to complete for sources stronger than 1.5~Jy 
at 23~GHz. 
\end{abstract}


\section{Introduction}
Large-area high-frequency radio surveys are time-consuming, and as a result relatively few large 
scale surveys have been carried out. Those that have been completed are either deep but covering small
areas, or shallow all-sky surveys. As a result, our knowledge of the high frequency 
($>10$~GHz) radio source population is poor.

In addition to the scientific benefits of studying the radio source population at high
frequencies, large scale surveys are useful for foreground subtraction for Cosmic Microwave
Background (CMB) experiments. 
Measurement of CMB anisotropies is limited by contamination from astronomical
foregrounds, both Galactic and extragalactic. To improve the efficiency of the 
component separation techniques, observations such as those of ESA's Planck mission are 
performed on a broad spectral region ranging from a few tenths to hundreds of GHz.
At frequencies up to $\sim100$~GHz, extragalactic radio sources are the major 
contaminants on angular scales smaller than 30 arcminutes \citep{dezotti05}, so 
identification of radio sources at high frequencies is critical.

The first blind radio survey above 8~GHz was carried out by \citet{taylor01} with the Ryle 
telescope at 15.2~GHz. The survey covered a 63 deg$^2$ region and detected 66 sources to a 
limiting flux density of 20 mJy. \citet{waldram03} extended the survey to 520 deg$^2$, 
detecting 465 sources to a flux density limit of 25\,mJy (the 9C survey). Both surveys found 
that the existence and flux density of sources at 15~GHz cannot be accurately predicted  
by extrapolation from lower frequency radio surveys such as the NRAO VLA Sky Survey (NVSS) 
at 1.4~GHz 
\citep{condon98}, further demonstrating the need for large scale high frequency surveys for 
population characterisation and source subtraction in CMB studies.

The Wilkinson Microwave Anisotropy Probe (WMAP) survey, which covers the whole sky at
23, 33, 41, 61 and 94~GHz \citep{bennett03}, is the first all-sky radio survey above 5~GHz.
The WMAP Point Source Catalogue constructed from the 5~yr maps contains 390 sources 
\citep{wright09}, compared to 323 and 208 sources found in the previous 3~yr and 1~yr 
maps respectively \citep{bennett03,hinshaw07}. Recently, \citet{massardi09} have 
detected 516 sources in the 5~yr WMAP maps by exploiting a combination of blind and
non-blind detection approaches. Section~\ref{s_wmap} presents a detailed comparison of
our AT20G results with the 5~yr WMAP point-source catalogue. 

Table \ref{t_surveys} summarises some earlier large-area high-frequency radio surveys,
showing the context in which the AT20G survey was designed.
A pilot survey for the AT20G at 18.5~GHz was carried out 
in 2002 and 2003 with the Australia Telescope Compact Array (ATCA) \citep{ricci04,sadler06} 
and detected 173 sources stronger than 100\,mJy in the declination range 
$-60$\degr to $-70$\degr.
\citet{ricci04} confirmed that ATCA (with custom hardware) had the capability to rapidly
survey the sky at high frequencies. 
The differential source counts for extragalactic
sources from the pilot survey  were found to be in good agreement with the 
\citet{waldram03} 15~GHz survey.

\begin{table*}
\centering
\caption{Comparison of the AT20G with other high-frequency radio surveys. 
Note that while the WMAP images cover the whole sky, regions at low Galactic 
latitude ($|b|<5\degr$) are excluded from the point-source 
search. A reanalysis of the WMAP survey by \citet{massardi09} resulted in a catalogue
of 516 sources above a limit of 695 mJy.\label{t_surveys}}
\begin{tabular}{llrrrll}\hline
Survey & Frequency & Sky area & Flux limit & N sources & Reference/s \\
       & (GHz)     & (deg$^2$) & (mJy)  & & \\
\hline
Ryle & 15                 & 60      & 20   & 66 & \citep{taylor01} \\
9C   & 15, 43             & 520     & 25   & 465 & \citep{waldram03} \\
9C   & 15                 & 115     & 10   &     & \citep{waldram09} \\
9C   & 15                 & 29      & 5.5   &     & \citep{waldram09} \\
WMAP & 23, 33, 41, 61, 94 & 32,177  & 1000 & 390 & \citep{bennett03,hinshaw07,wright09} \\
AT20G pilot & 18          & 1216    & 100  & 126 & \citep{ricci04} \\
AT20G & 5, 8, 20          & 20,086  & 40   & 5867 & \citep[This work;][]{massardi08} \\
\hline
\end{tabular}
\end{table*}

In this paper we present the full catalogue from the Australia Telescope 20~GHz
Survey (AT20G). The survey covers 20,086 square degrees (the complete Southern sky to declination
$0\degr$) to a limiting flux density of 40~mJy\,beam$^{-1}$. 
We followed up candidate sources detected in the survey at 20~GHz
and also have near-simultaneous follow-up observations at 5 and 8~GHz for AT20G sources 
south of declination $-15^\circ$.
An accompanying paper (Massardi et al., in preparation) will provide more 
detailed statistical analysis of the AT20G sample.
A subset of the 320 brightest ($S_{20} > 0.5$~Jy) extragalactic ($|b| > 1.5$\degr) AT20G sources 
were presented and discussed by \cite{massardi08}. 
The Galactic plane was included in our scanning survey but no follow-up
observations were carried out at $|b|<1.5\degr$, except for a blind survey of optically thick compact
HII regions \citep{murphy09}.

In Section~\ref{s_obs} we describe the survey and follow-up observations, and 
in Section~\ref{s_data} we describe the data reduction process. 
In Section~\ref{s_errors} we calculate the accuracy of our measured positions 
and flux densities. 
Section~\ref{s_cat} presents the source catalogue and defines its format, with 
the completeness and reliability of the catalogue discussed in Section~\ref{s_comp}.
As an additional investigation into the completeness, Section~\ref{s_wmap}
compares our catalogue with the 5~yr WMAP results. Finally, Section~\ref{s_prop} 
discusses some statistical properties of the sample and Section~\ref{s_conc} presents
our conclusions.

\section{Observations}\label{s_obs}
The key feature of the AT20G is a two-phase observing strategy. The first phase of our 
observations exploited the fast scanning capability of ATCA, using a wideband analogue 
correlator, to carry out a blind survey. Candidate sources from the scanning survey 
were then observed in the regular snapshot mode of the ATCA. 
The results from the scanning survey, which is complete
to a deeper level but has lower reliability, will be presented 
in a companion paper (Hancock et al., in preparation).

\subsection{Survey mode}\label{s_smode}
The first phase of our observations consisted of a raster of blind scans of the entire 
southern sky at 20~GHz, using the ATCA in fast scanning mode (in which it can 
achieve a speed of 15~degrees min$^{-1}$ in declination at the meridian).
The ATCA has low noise very wideband receiver\footnote{http://www.atnf.csiro.au/projects/mnrf1996/12mm\_details.html}
\citep{moorey08} which was used together with 
a custom analogue correlator. The correlator has 8~GHz bandwidth (Roberts, in preparation) and was
originally developed as part of the collaboration for the Taiwanese CMB experiment AMiBA \citep{lo01}.
We used it to take dual orthogonal polarization data from three of the six 22~m dishes of the ATCA. 
The lag-correlator measured 16~visibilities as a function of differential delay for each of the three 
antenna pairs used. 

Our custom correlator had no mechanism for allowing for geometrical delay as a 
function of the position in the sky, so the scans had to be performed along the meridian 
corresponding to zero delay for the East-West configuration ATCA. All the survey observations
used antennas 2, 3, and 4, situated on stations W102, W104, and W106, 
giving two 30.6~m ($2-3$ and $3-4$) baselines and a single 61.2~m ($2-4$) baseline. 
\changes{The shortest baseline of 30.6~m means the survey has reduced sensitivity to extended sources
(a 50 per cent reduction in amplitude for source size $>45\arcsec$). Larger sources will only be included 
in the survey if they have flux density in smaller angular size cores or hot spots above the survey limit.}

The scanning strategy consisted of sweeping sky regions 10$^\circ$ or 15$^\circ$ wide in 
declination and using Earth rotation to cover the full 24~hour right ascension
range in a zig-zag pattern. To achieve Nyquist sky coverage we used multiple zig-zag scanning
paths over multiple days -- each day the scanning path was shifted by half a beam width.
Along the scan a sample was collected every 54~ms (3~samples per beam), which resulted in
an rms noise of 12~mJy. Poor quality scans (due to weather or equipment error) were repeated,
so that the sky coverage is as uniform as possible.
The analogue correlator outputs for each set of 24~hour observations (interleaves) were combined 
together and calibrated to produce maps with an overall rms noise of $\simeq 10$~mJy.
Table~\ref{t_sobs} shows the observing schedule for each declination band in the scanning survey.
\begin{table}
\centering
\caption{Observation dates for the scanning observations for each
of the declination bands. Overlapping dates are due to patching data
being observed at the end of each years observations.}
\label{t_sobs}
\begin{tabular}{rr}
\hline
Declination Band & Observation Dates \\
\hline
$-30\degr\geq\delta\geq -40\degr$  & 11 Aug -- 31 Aug 2004 \\
$-40\degr\geq\delta\geq -50\degr$  & 20 Aug -- 31 Aug 2004 \\
$-50\degr\geq\delta\geq -60\degr$  &  9 Sep --  2 Oct 2005 \\
$-60\degr\geq\delta\geq -70\degr$  & 23 Sep --  2 Oct 2005 \\
$-70\degr\geq\delta\geq -80\degr$  & 16 Sep -- 20 Sep 2005 \\
$-80\degr\geq\delta\geq -90\degr$  & 20 \& 29 Sep 2005 \\
$-15\degr\geq\delta\geq -30\degr$  & 16 Aug --  3 Sep 2006 \\
$  0\degr\geq\delta\geq -15\degr$  & 23 Aug --  9 Sep 2007 \\
$-85\degr\geq\delta\geq -90\degr$  & 7 \& 9 Sep 2007 \\
\hline
\end{tabular}
\end{table}

The initial calibration of the interleaves was achieved by a daily transit observation of a 
nearby known calibrator. A second round of calibration was based on those sources from the 
ATCA calibrator catalogue\footnote{http://www.narrabri.atnf.csiro.au/calibrators} that fell 
within each of the interleave observations.
Typically there were about 10 such sources within each 24~hour observation. A third and final 
round of calibration was done using newly detected strong sources within the map itself. 

The delay steps in the analogue correlator were not precisely equal (as they are in a digital 
correlator) so the spectrum is not quite the same as the Fourier transform 
of the lags.  Hence standard synthesis techniques using discrete Fourier transforms could not be used 
for reliable source detection.  Furthermore the poor $(u,v)$ coverage of the survey (30 and 60~m 
EW spacings only) required a custom source detection program that worked in a
CLEAN-like fashion, detecting the strongest source, fitting a template
dirty beam and recording its location and flux estimate before moving
to the next source. Sources brighter than $5\sigma$ ($\sim50$~mJy\,beam$^{-1}$) were
scheduled for follow-up observations, as described in the next section.

\subsection{Follow-up mode}
\begin{figure}
\centering
\includegraphics[width=4cm]{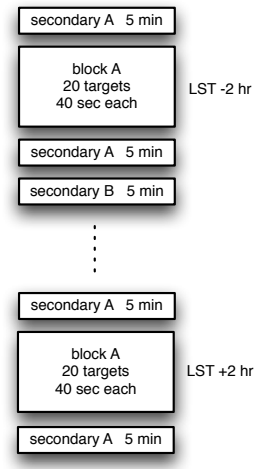}
\caption{Schematic of the follow-up observing schedule, showing blocks of target sources
enclosed by phase calibrators.}\label{f_sched}
\end{figure}
Each of the candidate sources identified in the first phase of observations was 
observed in the regular snapshot mode of the ATCA to confirm the detection and to measure 
an accurate position, flux density and polarisation. The flux densities from the initial 
scan survey are accurate to $\sim20$ per cent (see Hancock et al., in preparation) and so we expected a
fraction of the sources to fall below our specified survey detection threshold when re-observed.

The follow-up observations were scheduled to be as close as practical (typically within a 
few weeks) to the initial survey observations for a particular declination band, to 
reduce the effects of source variability on survey completeness.
Table \ref{t_obs} lists the details of each follow-up run, including the shortest antenna
spacing and resolution at each frequency.

The 20~GHz follow-up observations were performed using a hybrid array configuration 
(i.e., one in which there were North-South as well as East-West baselines) with the 
standard ATCA digital correlator.
We used two 128~MHz bands centred at $18\,752$~MHz and $21\,056$~MHz and two polarisations. 
During data processing the two bands were combined to form a single 256~MHz wide band 
centered at $19\,904$~MHz which is the reference frequency for our 20~GHz observations.

Candidate sources were prioritised for follow-up observations in order of decreasing flux
density to minimise problems with sample completeness. The sources were scheduled
in blocks of $\sim 20$ targets, with a nearby secondary calibrator observed for $\sim 5$~minutes 
at the beginning and end of each block. The sources within each block were observed 
in the fast mosaicing mode of the ATCA to reduce the slew time between pointings. We aimed 
to get two or more 40~second cuts for each source, at different hour angles, so as to obtain 
reasonable ($u,v$) plane coverage. Fig.~\ref{f_sched} shows a schematic of the observing 
schedule. Using this method up to 500 candidate sources were observed each day.

For declinations $<-15\degr$ each 20~GHz observing run was followed by lower frequency
observations of the same sets of sources. In these runs we used an East-West extended 
array configuration, with two 128~MHz bands centred at $4\,800$~MHz and $8\,640$~MHz. For 
the rest of this paper, these frequencies will be referred to as `5' and `8'~GHz. The lower 
frequency observations were conducted within a couple of weeks of the 20~GHz observations, 
so as to obtain near simultaneous spectral data for our sample. Sources in the most northern
declination band ($\delta>-15\degr$) were not observed at 5 and 8~GHz because of the poor 
($u,v$) coverage of EW arrays near the equator.

We completed several extra observing runs, at both 20 and 5/8~GHz for 
the purposes of replacing bad quality data from previous runs. 
Table \ref{t_obs} lists the details of our observing runs, with the first 
column showing the matching high frequency-lower frequency epochs.
The primary beam FWHM of the ATCA is $2.4$,  $5.5$, and $9.9$~arcmin at 20, 8,
and 5~GHz, respectively.

\begin{table*}
\centering
\caption{Follow-up observations at 20~GHz (C),
to observe them at 5 and 8~GHz (O) or to repeat previous bad quality observations
(R). 
\label{t_obs}}
\begin{tabular}{@{}ccrrrrrrrr@{}}
\hline
Epoch & \multicolumn{1}{c}{Declination} & Freq. 1       & Freq. 2 & \multicolumn{1}{c}{Array} & \multicolumn{1}{c}{Shortest} & \multicolumn{1}{c}{Beam 1$^a$} & \multicolumn{1}{c}{Beam 2} & \multicolumn{1}{c}{Dates} & Reason \\
      &  \multicolumn{1}{c}{range}     & \multicolumn{2}{c}{MHz} & \multicolumn{1}{c}{Config.} & \multicolumn{1}{c}{spacing} & \multicolumn{1}{c}{arcsec} & 
      \multicolumn{1}{c}{arcsec} & \\
\hline
1&$-50\degr$ to $-30\degr$  &  18752 & 21056  &   H214 & 80~m  &$10.8\times10.8$    &            & 21 Oct -- 27 Oct 2004 & C   \\
1&$-50\degr$ to $-30\degr$  &  4800 & 8640    &   1.5C & 77~m  &$8.3\times12.8$ & $4.6\times7.13$ & 04 Nov -- 08 Nov 2004 & O   \\
2&$-90\degr$ to $-50\degr$  &  18752 & 21056  &   H168 & 61~m  &$13.9\times13.9$ &               & 27 Oct -- 31 Oct 2005 & C   \\
2&$-90\degr$ to $-50\degr$  &  4800 & 8640    &   1.5C & 77~m  &$8.3\times8.8$ & $4.6\times4.9$   & 12 Nov -- 15 Nov 2005 & O   \\
3&$-90\degr$ to $-30\degr$  &  18752 & 21056  &   H214 & 80~m  &$10.8\times10.8$ &                & 29 Apr -- 03 May 2006 & R   \\
3&$-90\degr$ to $-30\degr$  &  4800 & 8640    &   1.5D & 107~m &$8.3\times9.5$ & $4.6\times5.3$  & 19 Jun -- 23 Jun 2006 & R,O \\
4&$-30\degr$ to $-15\degr$  &  18752 & 21056  &   H214 & 80~m  &$2.0\times5.1$  &                & 14 Oct -- 17 Oct 2006 & C   \\
4&$-30\degr$ to $-15\degr$  &  4800 & 8640    &   1.5B & 30~m  &$8.3\times21.1$ &$4.6\times11.7$ & 09 Nov -- 12 Nov 2006 & O   \\
5&$-90\degr$ to $-15\degr$  &  18752 & 21056  &   H214 & 80~m  & $10.8\times10.8$ &  & 11 May -- 16 May 2007 & R   \\
5&$-90\degr$ to $-15\degr$  &  4800 & 8640    &   1.5C & 80~m  &$8.3\times21.1$ &$4.6\times11.7$  & 04 May -- 10 May 2007 & R,O \\
6&$-90\degr$ to $0\degr$  & 18752 & 21056   &   H214 & 80~m  & $10.8\times10.8$               &                & 26 Oct -- 30 Oct 2007 & C,R \\
7&$-15\degr$ to $0\degr$ & 18752 & 21056   &     H75 & 31~m  & $33.9\times33.9$ &              & 22 Aug -- 26 Aug 2008 & R \\
\hline
\end{tabular}
\flushleft
\(^{\rm{a}}\) Note that the beam size changes with declination, and so these are estimates based
on the typical declination of sources observed in a particular epoch.
\end{table*}

\subsection{Additional Observations}
In addition to the main survey described in this paper, we have completed several
complementary sets of 
observations which will be presented in accompanying papers. In October 2006 
we observed a sub-sample of the bright sources to obtain high sensitivity 
polarisation measurements (Burke-Spolaor et al. in preparation). These observations 
used the most compact configuration of the ATCA (H75), which provides the best coverage
of a range of short spacings, and longer integration times than 
for the main survey, so that higher quality images could be made.

Nine highly extended sources were selected from low frequency catalogues -- PMN at 
4.85~GHz \citep{griffith93} and SUMSS at 843 MHz \citep{mauch03} -- and observed 
in mosaic mode to improve the flux density estimation at 20~GHz \citep{burke09}.
Data for seven of these sources have been incorporated into our catalogue in order 
to avoid flux density underestimation due to resolution effects. However, since 
we do not have equivalent 5 and 8~GHz observations of these objects we cannot 
use these data for analyses of radio spectra.

During the AT20G follow-up survey, the 6~km antenna of the ATCA was operational 
for most of the observing time, however the data from the five much longer baselines 
could not be easily included in our standard processing pipeline.    
As a separate program, we estimated the 6~km visibility by taking the ratio of the 
scalar amplitude at the 6~km baseline to the scalar amplitude at shorter baselines.  
This removes the effect of atmospheric decorrelation and avoids the need to phase 
calibrate the long baselines.  Approximately 90 per cent of the AT20G sources have 6~km 
data available.  These 6~km visibilities allows us to identify sources that appear 
point-like in the compact configuration, but extended on the 0.3\,arcsec scale corresponding 
to the 6~km baselines.  The full analysis and results from this high resolution 
follow-up will be presented in Chhetri et al. (in preparation).   An immediate application 
of this data was the identification of extended flat spectrum sources which are either 
thermal, or compact non-thermal sources extended by gravitational lensing.  
Gravitational lens candidates are being observed with ATCA 6~km configurations at 7~mm.   
The thermal sources identified by this procedure are Galactic objects, mainly planetary 
nebulae.  They are flagged as Galactic in the main catalogue -- see Section~\ref{s_galactic}.

\subsection{Galactic Plane Follow-up}
Neither the scanning survey nor the follow-up observations were well suited to 
imaging diffuse Galactic sources. We decided to exclude sources within 
$|b|<1.5\degr$ from our main follow-up survey and instead follow up selected 
subsamples of these sources with targeted observing runs. The first of these 
was the follow-up of a sample of ultra- and hyper-compact HII regions, selected on 
the basis of their rising spectral index between 843~MHz (from the 2nd Epoch 
Molonglo Galactic Plane Survey \citep{murphy07}) and 20~GHz ($\alpha(0.843,20)>0.1$). 
Further information is given in \citet{murphy09}.

\subsection{Variability Sample}\label{s_var}
Many large-area radio continuum surveys have been conducted at frequencies of 
1.4~GHz or below, where the long-term variability of most radio sources is low.  
As a result, the source catalogues from these surveys can continue to be used with 
a high level of confidence for many years after the survey was made.  
This is not necessarily true for surveys at frequencies above 5~GHz, where the source 
population is increasingly dominated by flat-spectrum radio sources which are 
expected to be variable \citep[e.g.][]{condon89}.

To study the variability of the 20~GHz radio-source population on timescales of years, 
we re-observed a sample of 170 sources (at declination $-60\degr$ to $-70\degr$ 
and with 20~GHz flux densities above 100~mJy) at several epochs over 
the course of the AT20G survey.  All the sources in this variability sample were originally 
detected in the AT20G Pilot Survey in 2002--2003 \citep{ricci04}, giving up to five epochs of 
observation for some of these sources.

The early (2002--2004) 20~GHz data for the variability sample are discussed by 
\citet{sadler06}, who conclude that the general level of variability in sources selected 
at 20~GHz is relatively low on timescales of $1-2$ years, with a median variability index 
of 6.9 per cent at 20~GHz over a one-year time interval.  
We therefore expect the AT20G catalogue presented here to be reasonably stable, 
in the sense that if we were to reobserve the survey area on timescales of a few years,
would expect the new source catalogue to contain most of the same sources as the old one. 

The AT20G catalogue presented in this paper only lists a single epoch for sources 
in the variability sample (typically the 2004 observation, except in cases of 
poor-quality data).  A separate paper (Sadler et al., in preparation) will present 
a full catalogue and detailed analysis of the AT20G variability sample.

\section{Data Reduction}\label{s_data}
We developed a fully automated custom analysis pipeline to edit, calibrate, and 
image the data from the follow-up observations. This included a suite of quality control
routines to ensure consistent data quality in the final catalogue.
The software was developed using the scripting language 
Python, and the data reduction was done with the aperture synthesis reduction 
package Miriad \citep{sault95}.

\begin{figure}
\begin{center}
\includegraphics[width=8cm]{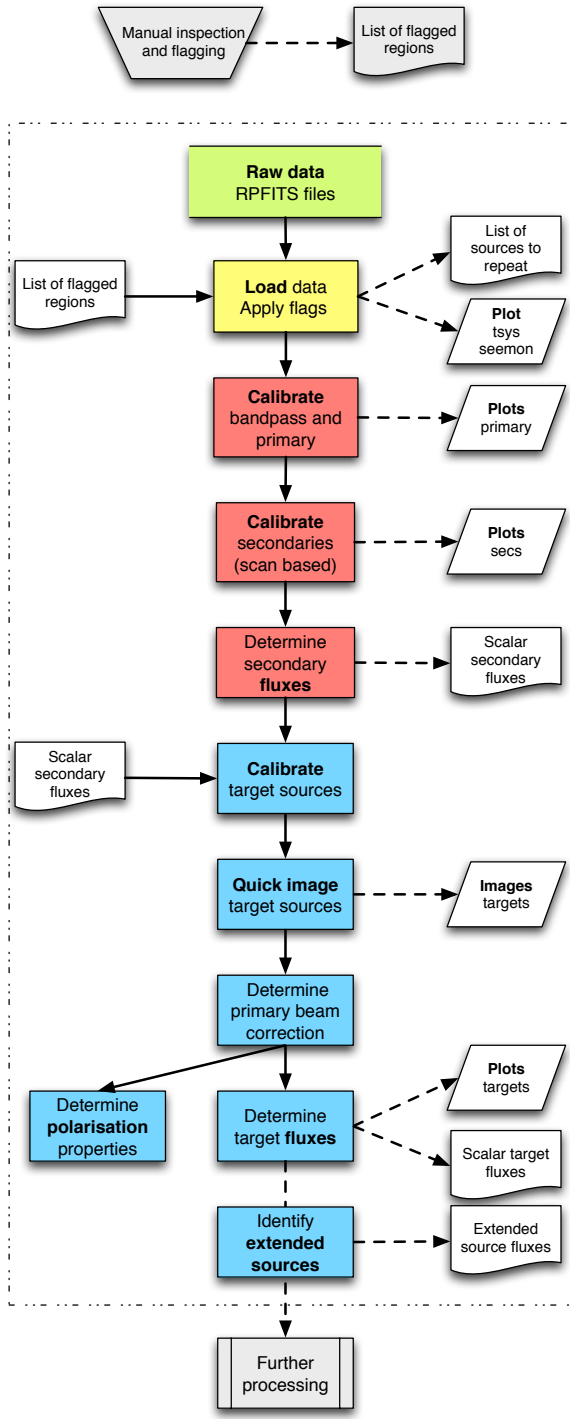}
\caption{An overview of the AT20G custom data reduction process. See Fig~\ref{f_cal} for
more detail about calibration.\label{f_pipeline}}
\end{center}
\end{figure}

The data reduction was followed by a cataloguing stage in which the best quality 
sources were selected for the final catalogue. This stage also included manual 
quality control, in which all sources included in the final catalogue were inspected 
by several AT20G team members (EM, EMS, JAE, RDE, TM). In the rest of this section we 
describe the details of the data reduction, quality control and cataloguing. An overview 
of the process is given in Fig.~\ref{f_pipeline}.

\subsection{Flagging poor quality data}
Weather conditions can seriously affect the quality of the high-frequency data. 
Attenuation of the signal 
by atmospheric water vapour can decrease the sensitivity of the observations, and atmospheric 
turbulence can produce phase fluctuations that may produce visibility amplitude decorrelation. 
Hence data collected in periods of bad weather were removed before further processing. 
In particular, calibrator data must be of high quality otherwise it introduces errors in the 
calibration solutions that affect the whole dataset.

An atmospheric seeing monitoring system operates at the ATCA site. This measures the differential phase 
variations in a geostationary satellite signal caused by tropospheric water vapour fluctuations 
\citep[as described in][]{middelberg06}. We used this data, in conjunction with the system 
temperature measurements from the antenna receivers  (to estimate tropospheric 
opacity) to develop semi-automatic flagging criteria. Specifically, we discarded data from 
all the periods in which there was decorrelation greater than 10~per cent. In cases where a 
calibrator was excluded, the block of target sources associated with that calibrator was also 
excluded.

In a majority of epochs less than a few per cent of the data was flagged.
Very occasionally, bad weather required large blocks of data in the follow-up survey to be edited 
out. Most of the time we were able to reobserve these blocks in clean-up runs. However, in the declination 
band $-15\degr$ to $0\degr$ there are several regions between 14~hr and 20~hr in right ascension
which are still incomplete.

\subsection{Calibration}
Primary flux calibration and bandpass calibration were done
in the standard way using PKS~B1934$-$638, with the assumed fluxes shown in Table~\ref{t_primary}. 
More information about the ATCA flux scale at 20~GHz is given in \citet{sault03}.
\begin{table}
\centering
\caption{Assumed fluxes for the primary calibrator PKS~B1934$-$638.}
\begin{tabular}{rr}\hline
Frequency & Flux \\
(MHz) & (Jy) \\
\hline
4800 & 5.83 \\
8640 & 2.84 \\
18752 & 1.04 \\
21056 & 0.88 \\
\hline
\end{tabular}
\label{t_primary}
\end{table}

\begin{figure}
\includegraphics[width=\columnwidth]{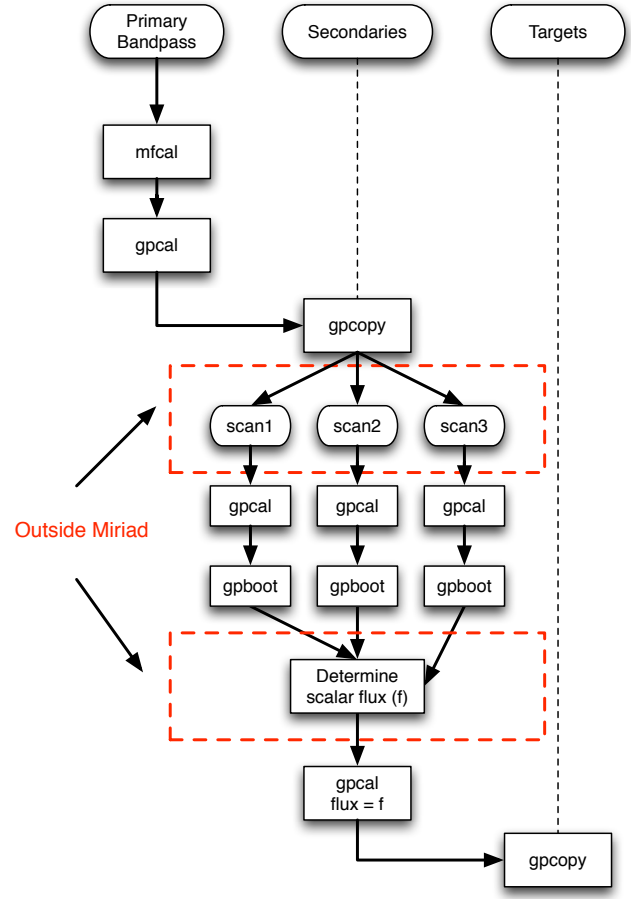}
\caption{The custom calibration process developed for the follow-up data reduction.}\label{f_cal}
\end{figure}

For the secondary flux calibration we followed a non-standard procedure which is 
summarised in Fig.~\ref{f_cal}. In each epoch of our observations we typically observed 
around $\sim 50$ secondary calibrators. To calculate an accurate flux density for each 
secondary calibrator we calculated the mean of the individual snapshot flux densities 
across the whole run, excluding snapshots which had a flux density greater than two 
standard deviations away from the mean. Each target source was then calibrated using 
the secondary calibrator associated with its observing block. 

Our observation strategy meant that we had insufficient data to determine the instrumental 
polarisation corrections from the secondary calibrators. Hence we calculated the leakage terms 
using the primary calibrator, PKS~B1934$-$638. The linear polarisation of this calibrator 
is known to be not variable and less than 0.2~per cent of the total source flux density 
at each of our observing frequencies. To determine the leakage terms it was assumed to be 
unpolarised. We adopted these leakage values for all the secondary calibrators, simultaneously 
calculating the time-dependent complex antenna gains, the residual xy phase differences (x and y
are the orthogonal linear polarisations), 
and the $Q$ and $U$ Stokes parameters of the calibrators. The polarisation calibration
was then applied to the target sources and $I$, $Q$, $U$ and $V$ Stokes 
parameters were determined for all of the target sources.

\subsection{Flux density measurements}\label{s_flux}

\subsubsection{Triple product fluxes}
We measured flux densities at all three frequencies using the triple product method 
implemented in the Miriad task {\sc calred}. The amplitude of triple product is the geometric 
average of the visibility amplitudes in a baseline closure triangle
\begin{equation}
A_{\rm TP}=\sqrt[3]{A_{1,2}\cdot A_{2,3}\cdot A_{3,1}}
\end{equation}
and its phase is the closure phase. Compared to measuring the flux densities from images, 
as is typical, this method of measuring flux densities is robust to the effects of phase 
decorrelation (Ricci et al., in preparation). This meant that we were able to recover 
flux density measurements for 143 of our target sources that could not be imaged due to 
poor weather conditions (these are marked as `poor' in the catalogue). 

\subsubsection{Extended source fluxes}
Sources that are partially resolved, or have multiple components, are not well-characterised
by the triple product flux measurement. We identified extended sources in our sample 
using several criteria based on the ratio of the triple product flux density to the 
flux density measured on the shortest baseline. We visually inspected the 20~GHz visibilities
and images for each of these sources, to confirm a sample of 337 extended sources.
Fig.~\ref{f_fim} shows an example of a typical AT20G extended source.

For the extended sources, we calculated the integrated flux densities at 5, 8 and 20~GHz
using the amplitude of the signal measured 
by the shortest baseline. Any source extended at 20~GHz was assumed to be extended at 
5 and 8~GHz. 
In most cases the shortest physical baseline used in the follow-up observations 
is either 60~m or 80~m (see Table~\ref{t_obs}), so our 20~GHz flux densities for sources 
larger than 20\,arcsec will still be underestimated using this method. Although improved 
algorithms could be used, the snapshot observations are inadequate to make reliable 
estimates. The extended sources will be discussed in more detail in a future paper. 
Note that the H75 array observations used for epoch 7 have closest spacing of 30~m so the 
short spacing flux estimate is good for sources up to about 1\,arcmin.

For extended sources with multiple components larger than 40\,arcsec, the shortest spacing can 
correspond to a minimum in the visibility and is not a useful estimate of total flux.  
For these sources we quote the triple product flux density that would correspond to the flux density 
of the dominant component and flag the flux as `poor' in the source catalogue.

Nine of the extremely extended sources were observed separately and we have 
used the flux densities determined in \citet{burke09} for seven of them. These are
discussed in Section~\ref{s_bandb}.

\subsection{Imaging}
\begin{figure*}
\centering
\includegraphics[width=3.3cm,angle=270]{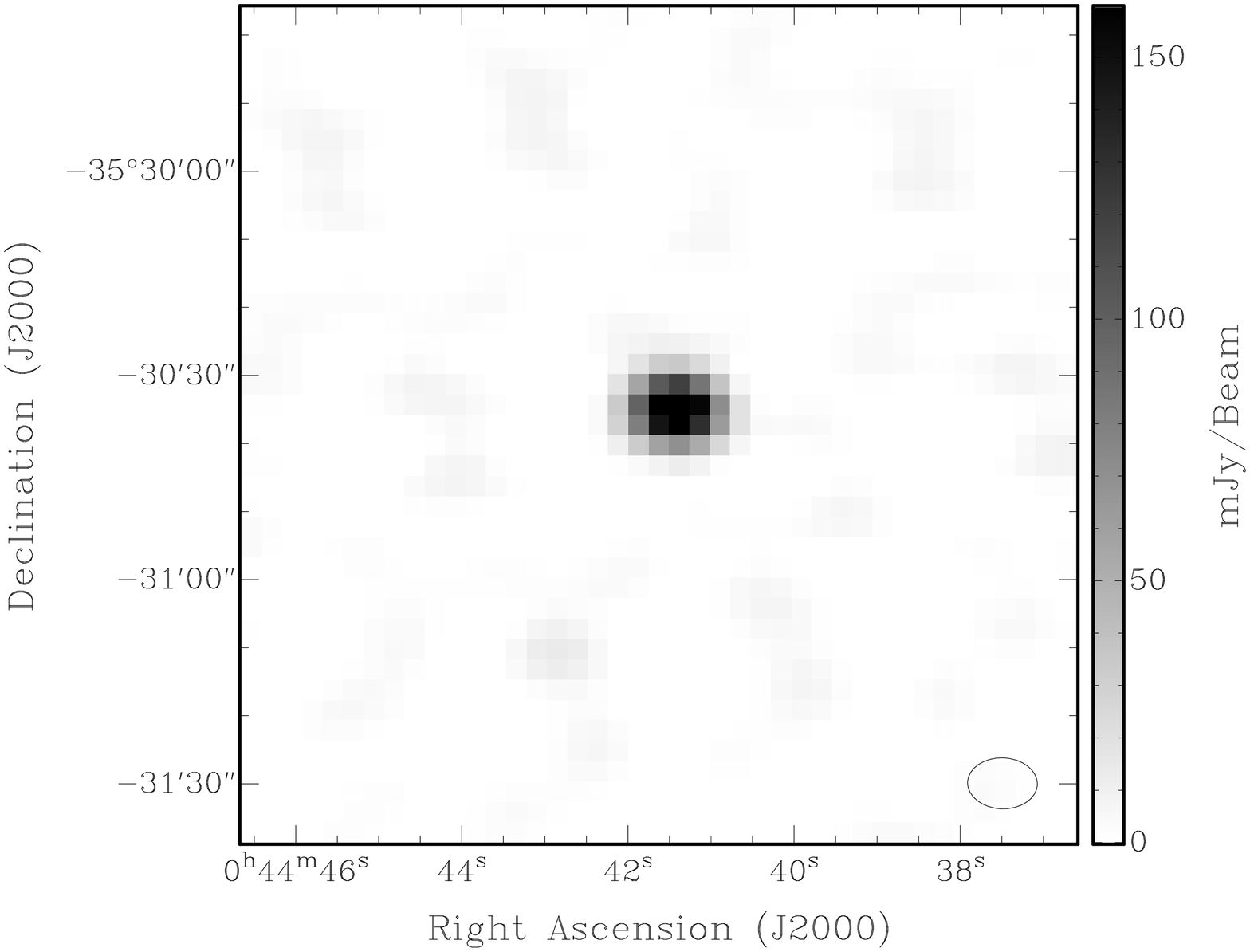}
\includegraphics[width=3.3cm,angle=270]{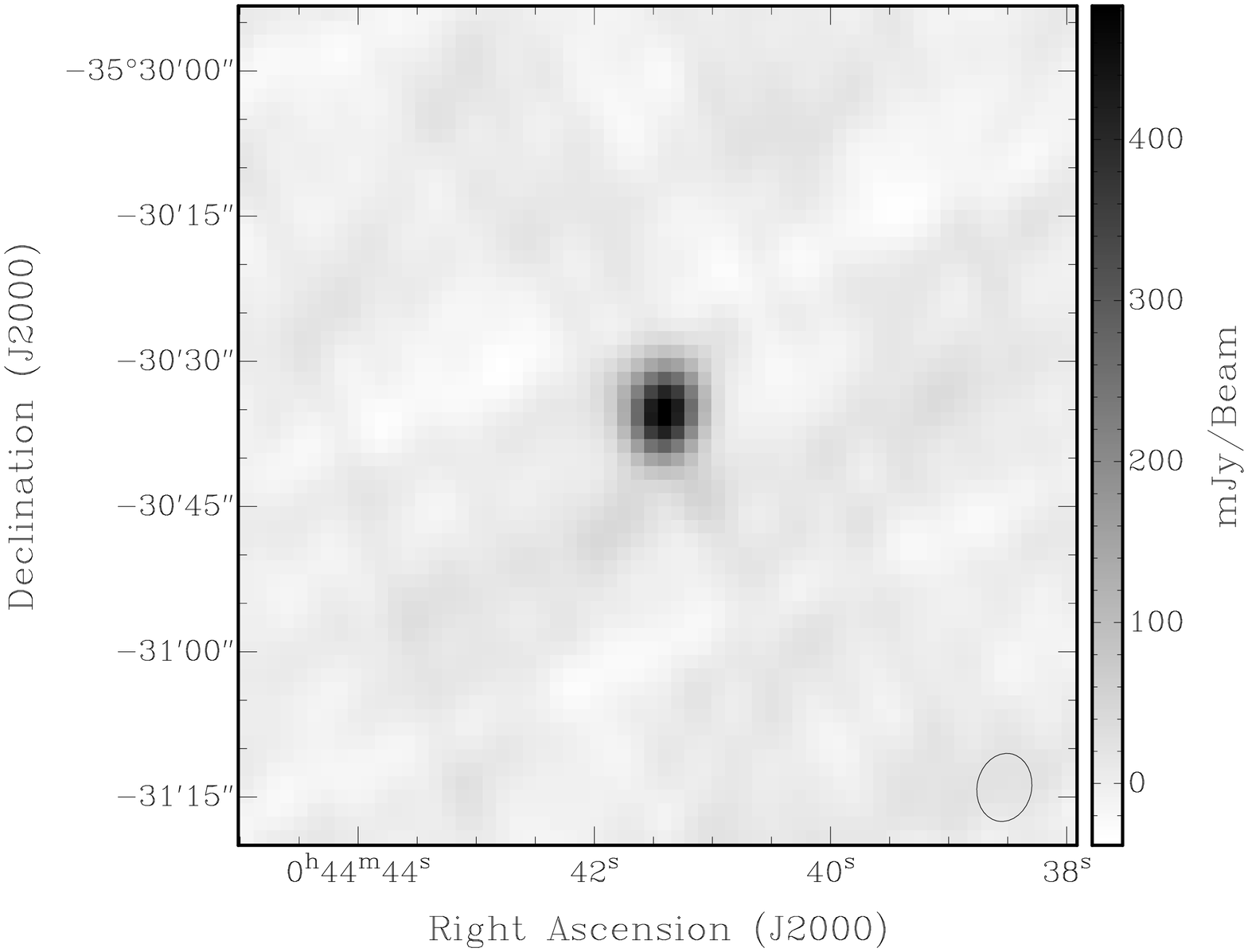}
\includegraphics[width=3.3cm,angle=270]{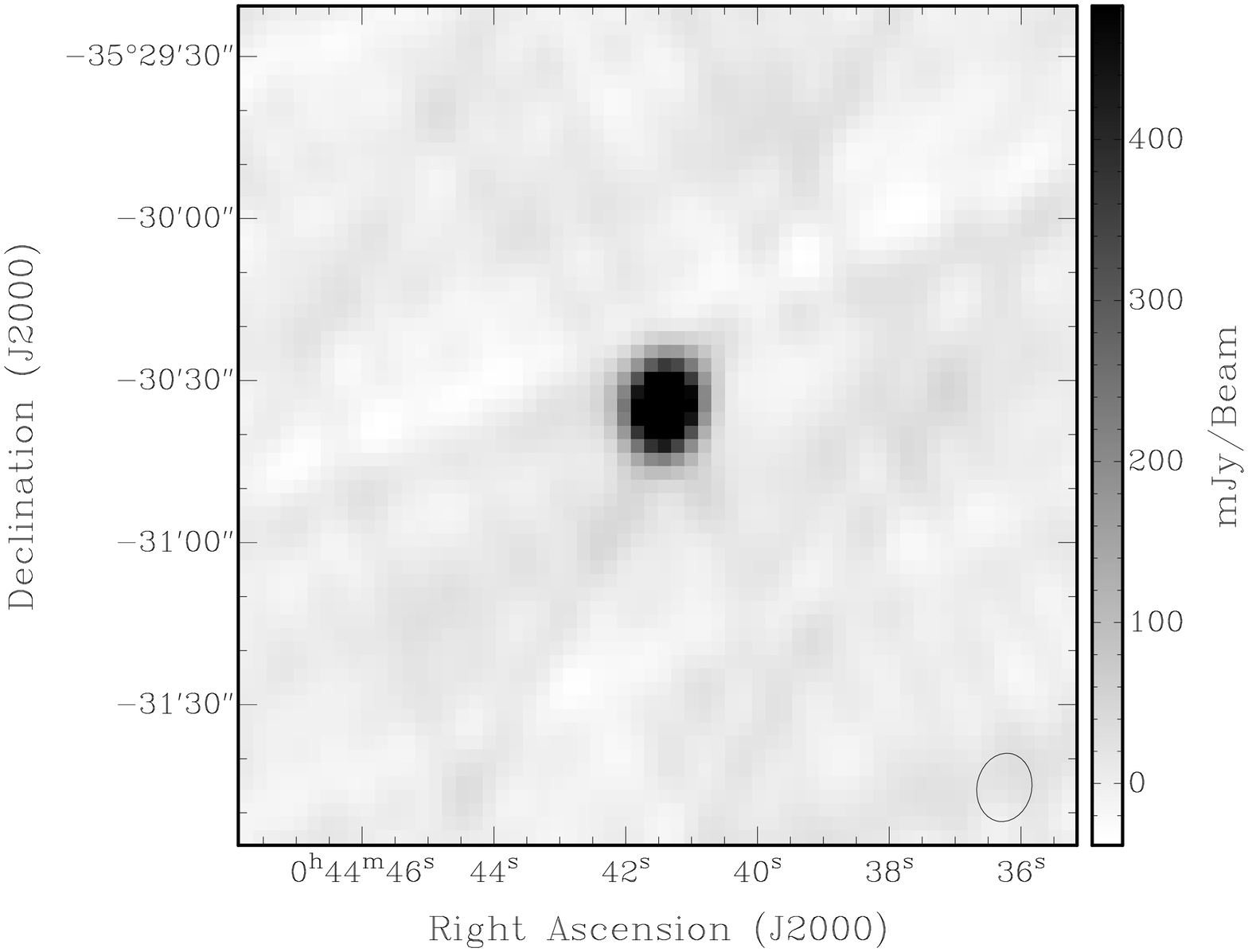}
\includegraphics[width=3.3cm,angle=270]{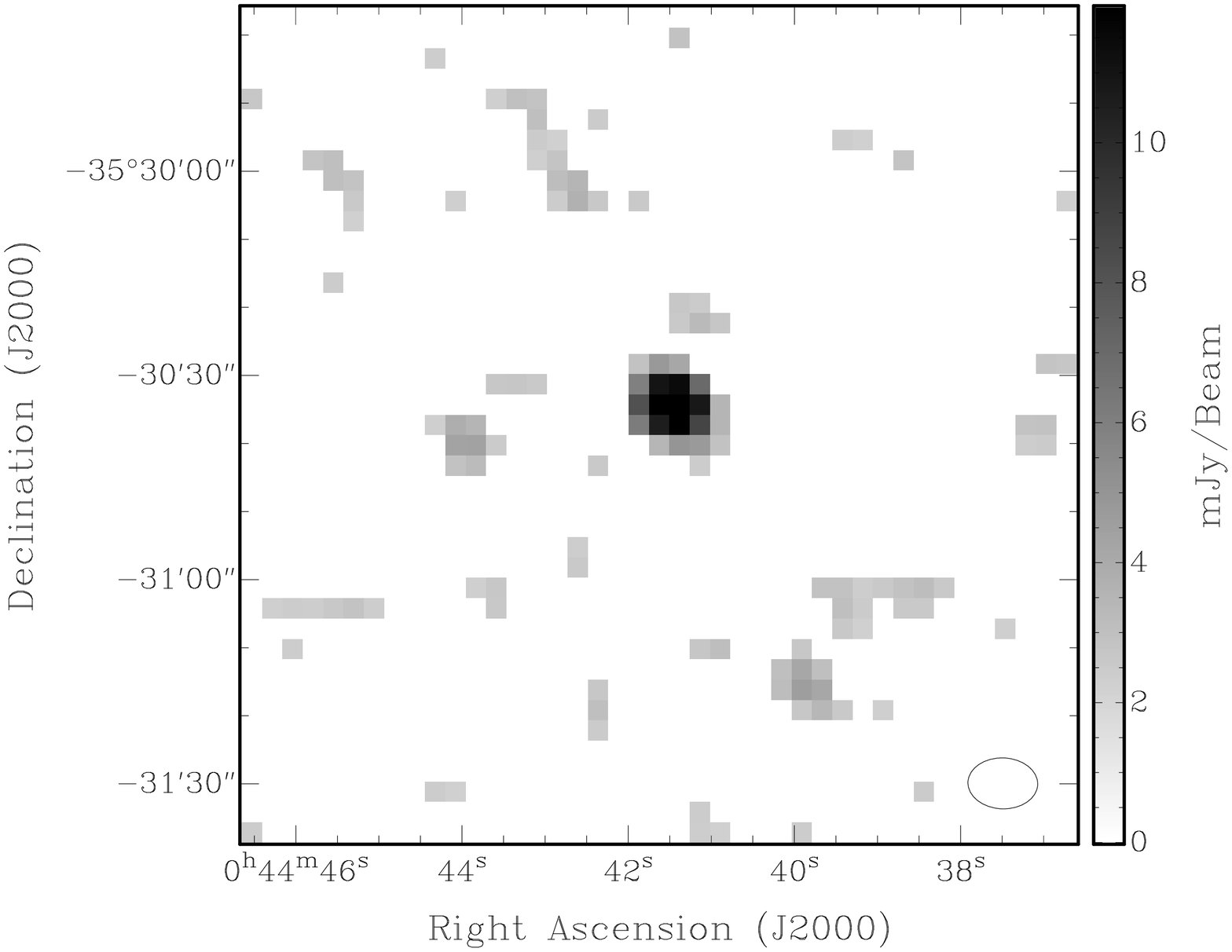} \\
\includegraphics[width=3.3cm,angle=270]{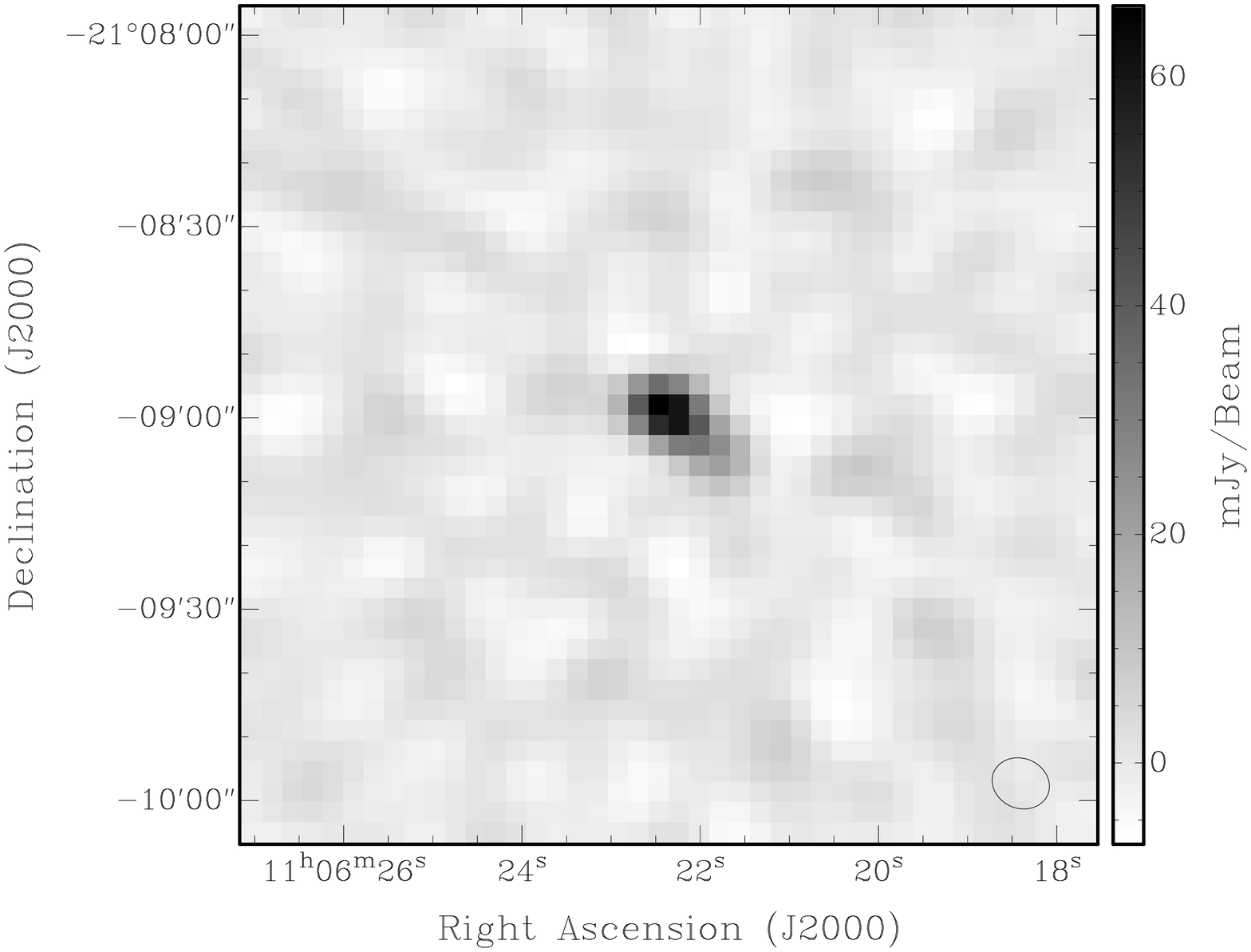}
\includegraphics[width=3.3cm,angle=270]{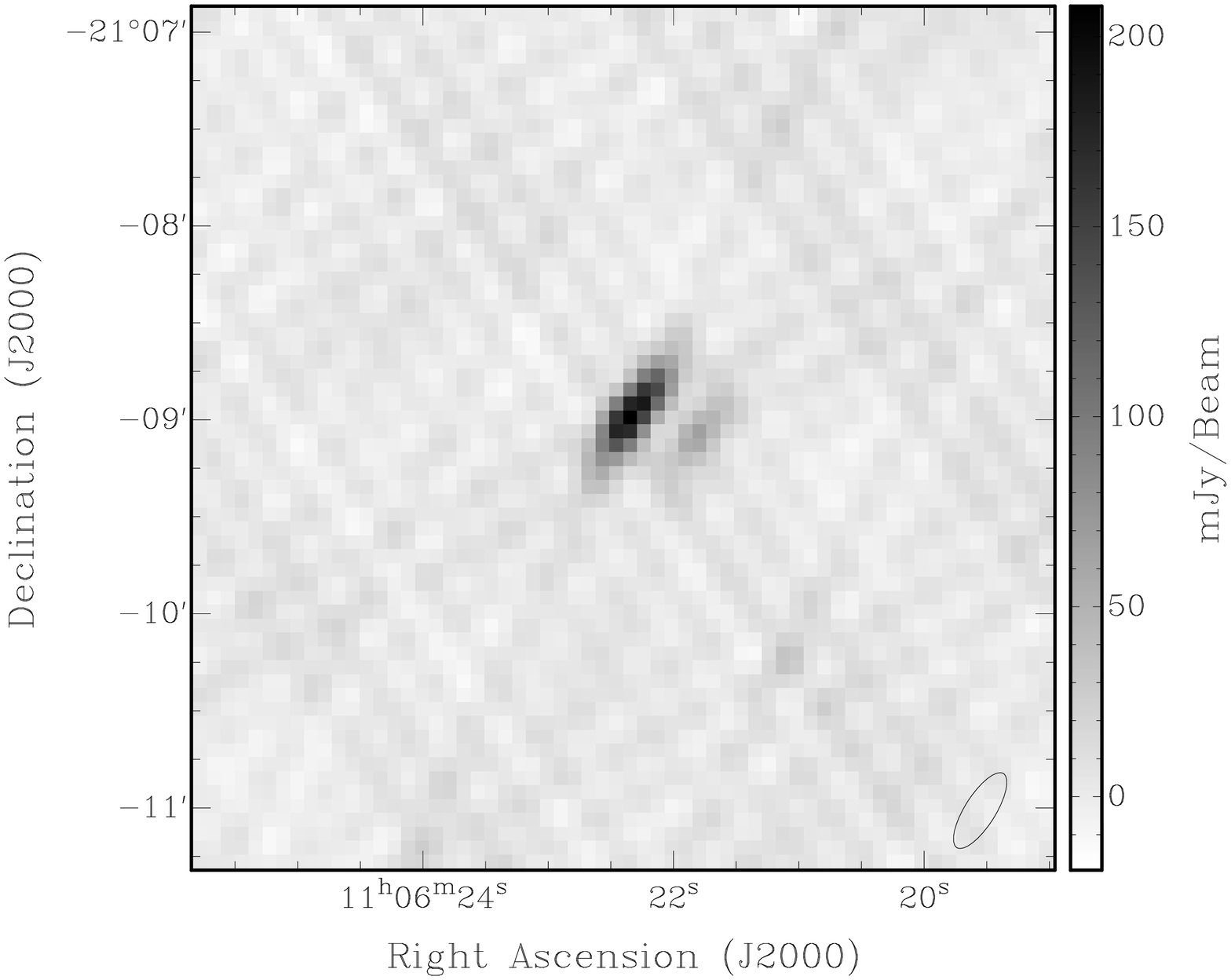}
\includegraphics[width=3.3cm,angle=270]{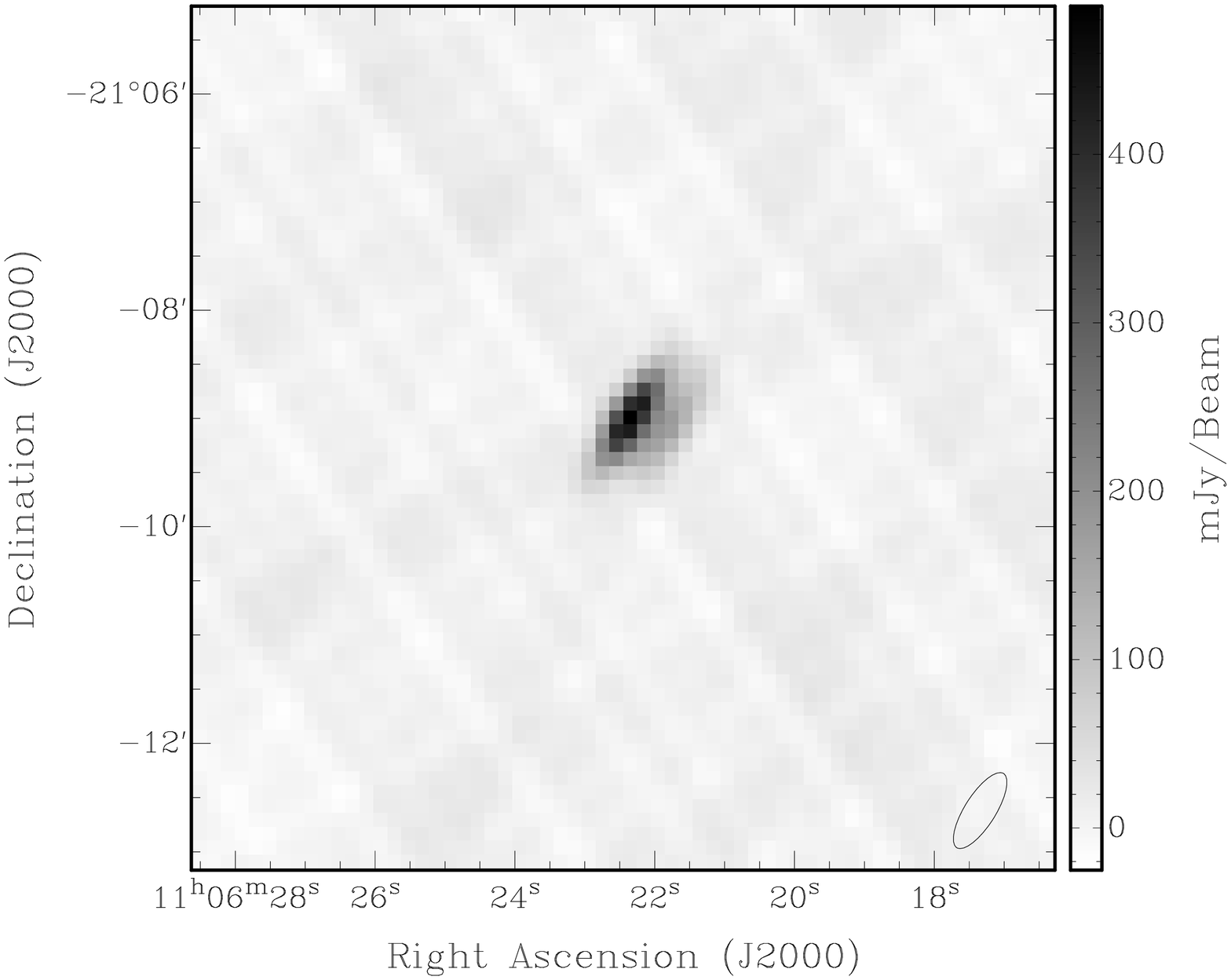}
\includegraphics[width=3.3cm,angle=270]{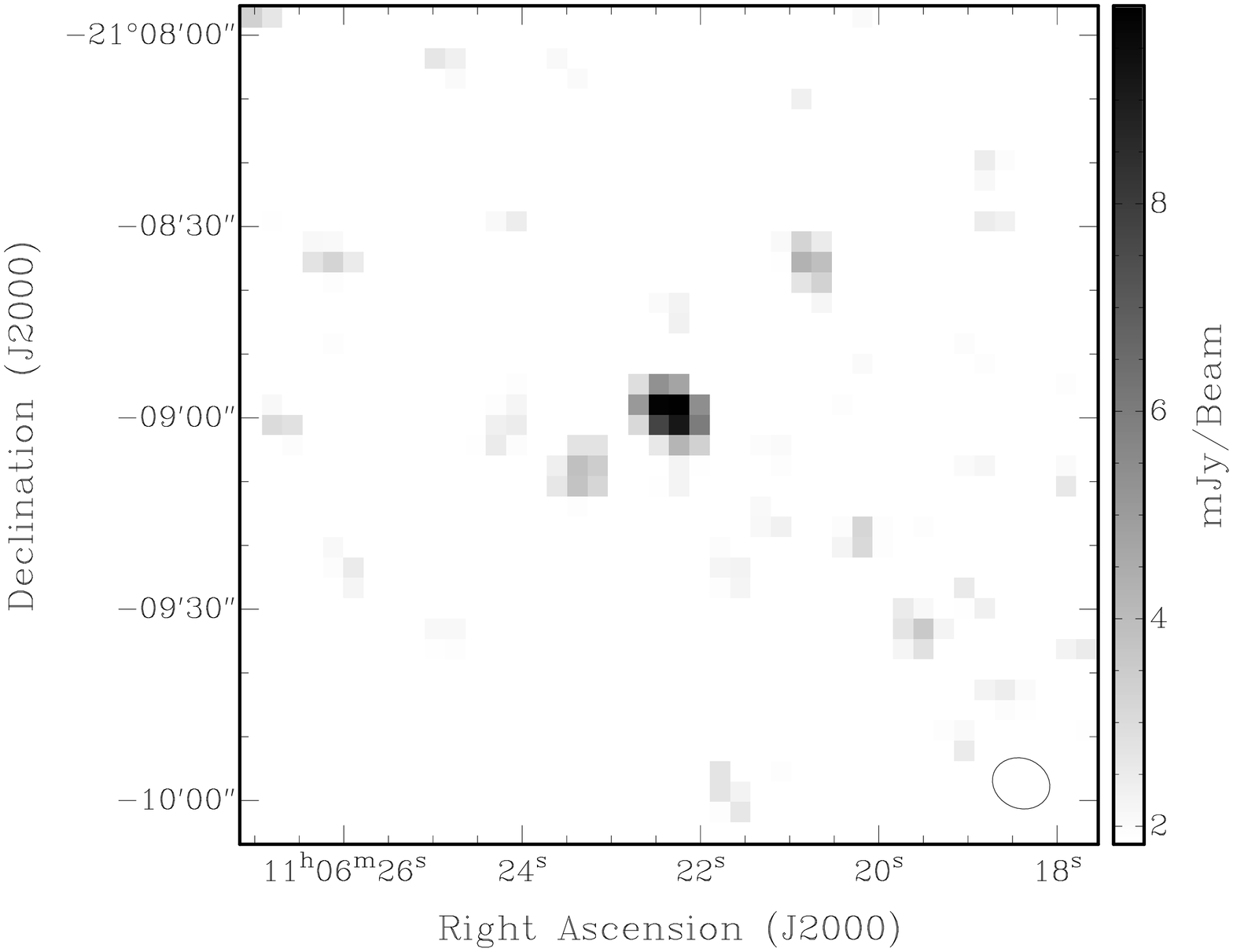}
\caption{Follow-up images for a typical AT20G point source, J004441$-$353034 (top) and
a typical extended source, J110622$-$210858 (bottom). From left to right they show the total 
intensity at 20~GHz, 8~GHz and 5~GHz, as well as the total polarization map. Note that 
each image has a different intensity scale.}\label{f_fim}
\end{figure*}

We imaged all sources using the standard Miriad process, and deconvolved them with a small
number of CLEAN iterations (typically 50). This was done primarily for the purpose of visual
inspection, since all flux densities were determined using the triple product as discussed in
the previous section. Fig.~\ref{f_fim} shows examples of a typical point source, 
AT20G J004441$-$353034, and a typical extended source, AT20G J110622$-$210858, at all three 
frequencies plus the total polarized intensity.

\subsection{Polarisation}\label{s_polmeas}
Polarization maps in the four Stokes parameters were created using the Miriad routine 
{\sc impol}. The total polarized intensity ($P$) for each source was measured by calculating 
the fractional polarization ($m$) in the image at the position of the source, and 
multiplying it by the flux density calculated using the triple product.
The error on the polarized flux was calculated using
\begin{equation}
P_{\rm err} = \sqrt{2} \frac{\sigma_{\rm n}}{\chi}
\end{equation}
where $\sigma_{\rm n}$ is the error associated with the noise, which was calculated from 
the rms noise measured in the Stokes V image, and $\chi$ is ratio of the flux density 
in the Stokes I image to the flux density as calculated from the triple product.
This corrects for phase decorrelation, which would affect both polarized and total flux 
density in the same way.

The polarised flux was measured at the position of the peak in the Stokes I image. This 
will be equivalent to the integrated polarization for the unresolved sources but will be 
neither the peak nor the integrated polarization for the extended sources. 
For the seven extremely extended sources discussed in Section~\ref{s_bandb} we have used the 
integrated polarization determined by \citet{burke09}.

We used the sources observed multiple times as part of the variability sample to 
investigate the reproducibility of our polarisation measurements (see Section~\ref{s_poln}). 
Based on these experiments we developed the following rules for defining a detection
\begin{itemize}
\item if $P \le 3P_{\rm err}$ then $P = 3P_{\rm err}$ as a limit
\item if $m \le 1\%$ then $P = 0.01S_{20}$ as a limit
\item if $P \le 6$ mJy then $P=6$ mJy as a limit
\end{itemize}
where $P$ is the total polarized intensity and $m$ is the fractional polarization.
If no polarization detection was made, then the limit on the polarized flux density is
the maximum of these three limits.

\subsection{Cataloguing}
All good quality observations were selected for the final source catalogue. 
In cases where a source was observed in multiple epochs (usually due to there being
a poor quality observation in the earlier epoch) we selected the best source based on
the data quality, the presence of near-simultaneous 5/8~GHz data and the offset of the 
source from the imaging pointing centre. For sources that were observed multiple times
as part of the variability study, we selected the first epoch of observation except in
cases where that observation was of poor quality.

\begin{figure}
\centering
\includegraphics[height=\columnwidth,angle=270]{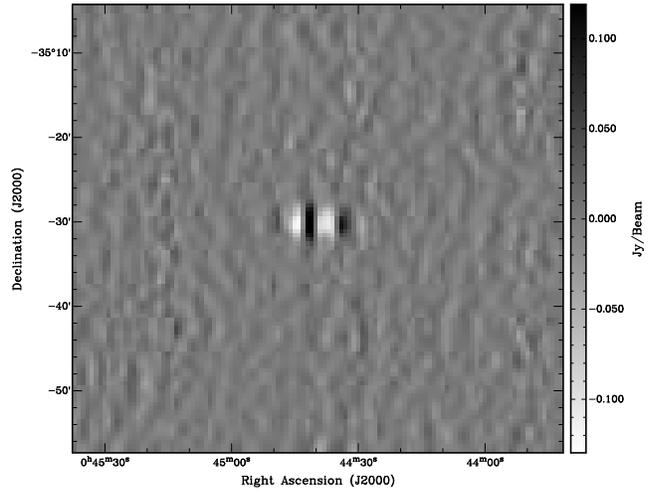}
\caption{Image of AT20G J004441$-$353034 (see Fig.~\ref{f_fim}), created from 
the scanning survey data.}\label{f_scanim}
\end{figure}

Due to the poor ($u,v$) coverage of the scanning survey, a small percentage of the 
sources scheduled for follow-up observations were in fact sidelobes of the main source. 
Fig.~\ref{f_scanim} shows a typical AT20G source as detected in the images created 
from the scanning survey. 
The scanning survey has only two different baselines (EW 30~m and 60~m) so the images have 
very high sidelobes. In some cases bright sidelobes were inadvertently followed-up. 
We excluded sidelobes automatically
by grouping all `duplicate' sources within a radius of 100\,arcsec (the effective beam size 
of the scanning survey) and filtering to keep only the strongest source in the group.
Cross-matching with low frequency surveys also helped to confirm these selections.

All sources were then inspected by several members of the AT20G team, using an online
annotation tool. We used this as a final quality control procedure, and to identify any
remaining sidelobes. We also used this process to identify extended sources that were 
not classified by our automatic selection criteria.

\section{Accuracy}\label{s_errors}

\subsection{Position uncertainties}
\changes{
The errors in position ($\sigma_\alpha$, $\sigma_\delta$) were calculated using
\begin{eqnarray}\label{e_pos}
\sigma^2_\alpha & = & \sigma^2_{\alpha, {\rm cal}} + \sigma^2_{\alpha, {\rm n}} \\
\sigma^2_\delta & = & \sigma^2_{\delta, {\rm cal}} + \sigma^2_{\delta, {\rm n}}
\end{eqnarray}
where $\sigma_{\rm cal}$ is the error associated with calibration, and $\sigma_{\rm n}$
is the error associated with the noise. }

\begin{figure}
\centering
\includegraphics[width=\columnwidth]{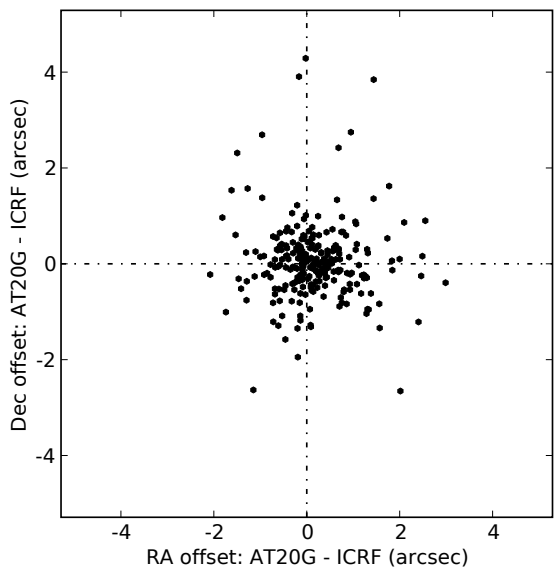}
\caption{Offset of AT20G 20~GHz positions from ICRF positions for the 251 ICRF calibrators 
in our catalogue.\label{f_icrf}}
\end{figure}

The calibration term was estimated by comparison with the the International Celestial Reference Frame 
(ICRF) defining calibrators \citep{fey04}. The VLBI-measured positions in the ICRF catalogue are 
accurate to a milliarcsecond or below, so any discrepancy between the positions 
of our target sources and ICRF positions can be assumed to be due to calibration 
positional errors in our sample.
There are 251 ICRF calibrators in our sample, and the positional offsets for these
are shown in Fig.~\ref{f_icrf}.
The mean offsets in right ascension and declination are
$\langle\Delta \alpha\rangle = 0.1$\,arcsec and 
$\langle\Delta \delta\rangle = 0.0$\,arcsec, showing we are in good agreement with the ICRF
positions. The values for rms scatter which we used as the calibration terms in 
Equation~\ref{e_pos} are $\sigma_{\alpha, {\rm cal}} = 0.8$\,arcsec and 
$\sigma_{\delta, {\rm cal}} = 0.9$\,arcsec.

\changes{The calibration term dominates over the noise for our data, are we assumed a common 
calibration term for all the target sources.}
From this we calculated the mean positional errors in right ascension and declination for 
the full sample, giving $\sigma_\alpha = 0.9$\,arcsec and $\sigma_\delta = 1.0$\,arcsec. Note that 
this method of calculating positional errors may underestimate the error for
extended sources.

\subsection{Flux density uncertainties}
We calculated the errors in the flux density measurements by adding in quadrature 
the error associated with calibration (gain error, $\sigma_{\rm gain}$) and the error 
associated with the noise level ($\sigma_{\rm n}$):
\begin{equation}
\sigma^2 = \sigma_{\rm gain}^2 S^2 +  \sigma_{\rm n}^2
\end{equation}

The gain error is a multiplicative term (i.e., it is proportional to the source flux 
density) and is a measure of the gain stability over time. We estimated 
$\sigma_{\rm gain}$ for each observational epoch and frequency from the scatter in
the visibility amplitudes of the calibrators in each observing run. The mean values for 
the gain errors were found to be of order a few per cent.
The noise term is an additive term related to the interferometer noise, which is 
proportional to the system temperature. Since no source has significant flux in Stokes $V$, 
the rms noise levels in the $V$ images have no gain error and were used as an estimate 
of $\sigma_{\rm n}$.
The error in the flux density for each source is given in the main catalogue 
(Table 4). Typically the error is $4-5$ per cent of the total flux density.

For extended sources the error was multiplied by the square root of the number of 
baselines $n_{\rm base}$ (normally 10 for our 5-antenna follow-up arrays) to correct 
for the fact that the flux densities for these sources are estimated using only one (the
shortest) baseline instead of $n_{\rm base}$.

\begin{figure}
\centering
\includegraphics[width=\columnwidth]{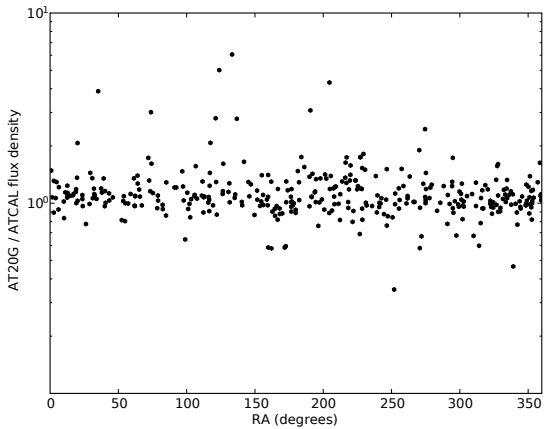}
\caption{The ratio of AT20G 20~GHz flux density to Australia Telescope Calibrator Catalogue
flux density, for the 362 AT calibrators in our catalogue. The AT calibrator flux density 
closest in time to our AT20G observations was chosen.\label{f_atcals}}
\end{figure}

Many of the sources in the Australia Telescope calibrator 
catalogue were observed as target sources as part of the AT20G program. 
We used these sources to check the integrity of our observing and data 
reduction procedures.
There are 362 AT calibrators in the final AT20G catalogue. For each of these
we extracted the AT calibrator catalogue flux density measurement that was 
closest in time to our AT20G observations of that source.
Fig.~\ref{f_atcals} shows the AT calibrator catalogue to AT20G flux density ratio 
for these sources. The median AT20G/AT flux density ratio is $1.06\pm0.03$. All of the outliers 
were found to be either extended or highly variable sources. 

A small number of our target sources are included in the VLA Calibrator 
Manual\footnote{\tt http://www.vla.nrao.edu/astro/calib/manual}. We used these
as an additional check on our flux density measurements, particularly for sources
near the equator.
The VLA calibrator data is only available up to the year 2000. This means 
there is a significance difference in the times that these measurements were made. 
The scatter we see in our sample is quite high, due to variability of the sources
over this 8~year period. The median AT20G/VLA flux density ratio for these sources 
is 1.08$\pm0.10$.

\changes{We have explored the reasons for the possible flux scale differences between 
AT20G, and ATCA and VLA calibrators. However, without simultaneous measurements on both
telescopes it is difficult to obtained a conclusive result. We see no evidence that our
catalogue flux densities are not on the scale set by the assumed flux of 1934$-$638. 
We note that our fluxes are consistent with the WMAP flux scale (see Section~\ref{s_wmap})
and strongly recommend a future program of simultaneous measurements of calibrator sources 
to tie the flux scales in the north and south hemispheres, and between the major radio observatories.}

\section{Source Catalogue}\label{s_cat}
The primary AT20G source catalogue gives the flux density and polarisation measurements
at 20, 8 and 5~GHz, as well as the epoch of observation and quality flags. 
Table~7 shows the first 30 sources in the catalogue, the full catalogue is included in
the Supporting Information for this paper, and available
online through Vizier\footnote{http://vizier.u-strasbg.fr}. The columns are:
\begin{description}
    \item[(1)] AT20G source name.
    \item[(2--3)] Right ascension and declination (J2000). The mean errors 
    ($\sigma_\alpha=0.9$\,arcsec, $\sigma_\delta=1.0$\,arcsec) are derived in Section~\ref{s_errors}.
    \item[(4--5)] Flux density at 20~GHz and error in mJy.
    \item[(6--7)] Flux density at 8~GHz and error in mJy.
    \item[(8--9)] Flux density at 5~GHz and error in mJy.
    \item[(10)] Epoch of observation for the three frequencies (20, 8 and 5~GHz). \\
    \hspace*{3mm} The epoch codes are listed in column 1 of Table \ref{t_obs}.
    \item[(11)] Quality flag (see description below)
    \item[(12)] Other flags (see description below)
    \item[(13--14)] Polarized intensity at 20~GHz and error in mJy.
    \item[(15)] Fractional polarisation at 20~GHz.
    \item[(16)] Position angle at 20~GHz.
    \item[(17--18)] Polarized intensity at 8~GHz and error in mJy.
    \item[(19)] Fractional polarisation at 8~GHz.
    \item[(20)] Position angle at 8~GHz.
    \item[(21--22)] Polarized intensity at 5~GHz and error in mJy.
    \item[(23)] Fractional polarisation at 5~GHz.
    \item[(24)] Position angle at 5~GHz.
\end{description}
Column 11 contains a quality flag, either {\it g} (good, 5501 sources) or 
{\it p} (poor, 389 sources). A poor quality flag indicates that there was 
lower quality data in that observation, or that, in the case of an extended 
source, the triple product flux was used rather than the shortest baseline flux. 
Hence the flux density measurement for poor quality sources may not be as reliable. 

Column 12 contains other flags identifying the following source types or issues: 
\begin{description}
     \item[{\bf e}] Source is extended and the shortest baseline flux has been used (see Section~\ref{s_flux}).
     \item[{\bf h}] Source identified as a Galactic HII region (Section~\ref{s_galactic}).
     \item[{\bf p}] Source identified as a Galactic Planetary Nebula (Section~\ref{s_galactic}).
     \item[{\bf m}] Source identified as part of the Magellanic Clouds (Section~\ref{s_mc}).
     \item[{\bf l}] Source has no match in the low frequency surveys (NVSS and SUMSS) (Section~\ref{s_low}).
     \item[{\bf b}] Source is large and extended (see Table~\ref{t_bandb}). The data in the catalogue comes from the observations presented in \citet{burke09} and discussed in Section~\ref{s_bandb}.
\end{description}

AT20G catalogue sources should be referred to by their full IAU
designation \citep{lortet94}. These are of the form AT20G {\it JHHMMSS$-$DDMMSS} where
AT20G is the survey acronym, {\it J} specifies J2000.0 coordinate equinox,
{\it HHMMSS} are the hours, minutes and truncated seconds of right
ascension, $-$ is the sign of declination and {\it DDMMSS} are the degrees,
minutes and truncated seconds of declination.

\subsection{Bright extended sources}\label{s_bandb}
As discussed in \citet{massardi08}, a small number of highly extended 
sources were expected to have a 20~GHz flux density above our bright source 
sample cutoff of 0.5~Jy, but were undetected or had diminished levels of 
observed emission in the AT20G follow-up observations due to the sources' 
extent beyond our observing resolution and field of view ($\sim2.4$ arcminutes). 
These sources were identified using extrapolated flux densities from the 
5~GHz PMN \citep{griffith93} and the 0.843~MHz SUMSS \citep{mauch03} data. 
Integrated flux density and polarisation measurements for these sources (shown
in Table \ref{t_bandb}) were 
measured from mosaic observations done during the 2006 October polarisation 
follow-up run. Their flux densities have been included in our main catalogue 
and are flagged with a {\bf b}. The observations and data analysis are 
discussed fully in \citet{burke09}. The objects with mosaiced 
measurements are limited to sources south of $\delta = -30\degr$. 

\changes{We note that as discussed in Section~\ref{s_smode} our survey will be incomplete 
for sources larger than 45\arcsec and based on simple source count arguments 
many hundred sources larger than a few arc minutes will be missing.  Since non-thermal 
extended sources have steep radio spectra we suggest that low frequency catalogues 
such as NVSS or SUMSS rather than AT20G be used if complete samples of the extended sources 
are important.}

\begin{table}
\centering
\caption{Properties of 7 bright extended sources from \citet{burke09} that have been 
included in the AT20G source catalogue. Core and total integrated flux density (or limits) 
and errors are given in Jy.}\label{t_bandb}
\begin{tabular}{cccc}
\hline
AT20G Name & Source Name & $S_{core}$ (Jy) & $S_{18}$ (Jy) \\ 
\hline
AT20G J013357$-$362935 & PKS 0131$-$36 & 0.03$\pm$0.01 & $>$0.44 \\ 
AT20G J051949$-$454643 & Pictor A      & 1.32$\pm$0.04 & 6.32$\pm$0.11 \\ 
AT20G J132527$-$430104 & Centaurus A   & 5.98$\pm$0.17 & $>$28.35 \\ 
AT20G J134649$-$602430 & Centaurus B   & 5.02$\pm$0.06 & 8.89$\pm0.43$ \\ 
AT20G J161505$-$605427 & PKS 1610$-$60 & 0.14$\pm$0.05 & 2.11$\pm$0.04 \\ 
AT20G J215706$-$694123 & PKS 2153$-$69 & ---           & 3.40$\pm$0.21 \\ 
AT20G J235904$-$605503 & PKS 2356$-$61 & 0.09$\pm$0.03 & 1.64$\pm$0.05 \\ 
\hline
\end{tabular}  
\end{table} 

\subsection{Comparison with bright source sample}
Some bright sources ($S_{20} > 0.5$ Jy) in this catalogue have slightly different
parameters than were given in the Bright Source Sample (BSS) paper \citep{massardi08}.
There are 23 sources which have a flux density that differs by more than 10 per cent from
the BSS results. In most cases this is due to a different observational epoch being
selected for the source in the BSS than for in the current catalogue. The largest 
group of sources in this category are sources from the variability sample discussed in
Section~\ref{s_var}. In the selection of sources for this catalogue we gave preference where
possible to the epoch with simultaneous 5 and 8~GHz observations. 
However, in the BSS the best quality epoch at 20~GHz was chosen regardless of the 
availability of follow-up at other frequencies.

\subsection{Magellanic Clouds}\label{s_mc}

The AT20G survey area includes the Large and Small Magellanic Clouds, and we would 
expect to find some radio sources in these galaxies, which are effectively foreground 
sources for cosmological studies.
In earlier work using the AT20G pilot survey data, \citet{ricci04} and \citet{sadler06} 
simply removed $5^\circ\times5^\circ$ regions of 
sky around the LMC and SMC to avoid the problem of foreground 
contamination. 
However, since the compact, high-frequency LMC/SMC radio sources are 
likely to be interesting in their own right, we have now attempted to 
identify them individually in the final AT20G catalogue.  

We searched two areas of sky defined by the following (J2000) coordinates: \\

\begin{tabular}{lll}
SMC & 00:30 $<\alpha<$ 01:30 &  $-$71\degr $<\delta<$ $-$75\degr \\
LMC & 04:45 $<\alpha<$ 06:00 & $-$66\degr $<\delta<$ $-$72\degr \\
\end{tabular}
\linebreak

\noindent There are six AT20G sources in the SMC region and 21 in the LMC region. 
As with Galactic sources, some LMC and SMC radio sources may have complex, extended
emission which is not well imaged by the AT20G snapshot observations.  In such cases, 
the flux densities and positions listed in the AT20G catalogue should be regarded as 
no more than indicative. 
Hence one of the main reasons for identifying LMC/SMC sources 
is to have the option of excluding them from later analysis of the extragalactic 
AT20G sample. 

To determine whether each 20~GHz source was likely to be associated with
the LMC or SMC we searched the NASA Extragalactic Database 
(NED)\footnote{\tt http://nedwww.ipac.caltech.edu} for known LMC/SMC objects at
or near the AT20G position, and cross-matched the AT20G positions with
lower-frequency radio catalogues from \citet{filipovic95} and \citet{payne04},
who independently classified many of these sources as either LMC/SMC 
or background objects. We also checked for optical nebulosity of the sources by 
inspecting overlays of AT20G contours on optical images from SuperCOSMOS.

In most cases, it was possible to distinguish foreground LMC/SMC objects 
from background AGN with a high level of confidence. We classified 
14 of the 27 sources as background extragalactic objects and 13 
as LMC/SMC radio sources (listed in 
Table \ref{lmc_smc} and identified in the catalogue with a flag `{\it m}').  
As expected most of the LMC/SMC objects are HII regions, but the 
AT20G catalogue also includes two LMC pulsar wind nebulae, PSR~J0537$-$69
and PSR~B0540$-$69.3. 

\begin{table}
\centering
\caption{AT20G sources which are probably associated with objects in the Large and Small 
Magellanic Clouds. HII $=$ HII region, PWN $=$ pulsar wind nebula.  Objects in the 
LMC and SMC regions which we classified as background sources are also listed. }\label{lmc_smc}
\begin{tabular}{lll}
\hline
AT20G name & AT20G position (J2000) & Notes  \\ 
\hline
\multicolumn{3}{l}{\it SMC objects} \\
J012407$-$730904  & 01 24 07.92 $-$73 09 04.1 &  IRAS\,01228$-$7324, HII \\
J012930$-$733311  & 01 29 30.07 $-$73 33 11.3 &  IRAS\,01283$-$7349, HII \\
\multicolumn{3}{l}{\it LMC objects} \\
J045153$-$692329  & 04 51 53.25 $-$69 23 29.4 &  IRAS\,04521$-$6928, HII \\
J050950$-$685305  & 05 09 50.61 $-$68 53 05.6 &  Part of NGC\,1858, HII \\
J051317$-$692222  & 05 13 17.75 $-$69 22 22.6 &  Large HII region complex \\
J052212$-$675832  & 05 22 12.67 $-$67 58 32.9 &  Part of NGC\,1936, HII \\
J053745$-$691010  & 05 37 45.51 $-$69 10 10.0 &  PSR~J0537$-$69, PWN \\
J053845$-$690503  & 05 38 45.66 $-$69 05 03.1 &  Part of NGC\,2070, HII \\
J053937$-$694526  & 05 39 37.41 $-$69 45 26.4 &  Part of NGC\,2079, HII \\
J053945$-$693839  & 05 39 45.57 $-$69 38 39.2 &  NGC\,2080, HII \\
J054004$-$694438  & 05 40 04.76 $-$69 44 38.6 &  Part of NGC\,2079, HII \\
J054011$-$691953  & 05 40 11.09 $-$69 19 53.4 &  PSR\,B0540$-$69.3, PWN \\
J054024$-$694014  & 05 40 24.69 $-$69 40 14.5 &  IC\,2145, HII \\
\multicolumn{3}{l}{\it Background sources} \\
J004047$-$714559  & 00 40 47.90 $-$71 45 59.6 &  \\ 
J005611$-$710707  & 00 56 11.34 $-$71 07 07.0 &  \\ 
J011049$-$731428  & 01 10 49.61 $-$73 14 28.2 &  \\ 
J011132$-$730209  & 01 11 32.25 $-$73 02 09.9 & \\ 
J045551$-$690209  & 04 55 51.56 $-$69 02 09.5 &  \\ 
J045608$-$701433  & 04 56 08.67 $-$70 14 33.3 &   \\ 
J050551$-$695116  & 05 05 51.89 $-$69 51 16.5 &   \\ 
J051129$-$680618  & 05 11 29.40 $-$68 06 18.1 &   \\ 
J051222$-$673220  & 05 12 22.54 $-$67 32 20.5 &   \\ 
J051537$-$672128  & 05 15 37.36 $-$67 21 28.4 &   \\ 
J051832$-$693520  & 05 18 32.48 $-$69 35 20.6 &   \\ 
J052635$-$674909  & 05 26 35.05 $-$67 49 09.2 &   \\ 
J054317$-$662655  & 05 43 17.54 $-$66 26 55.8 &   \\ 
J054750$-$672801  & 05 47 50.02 $-$67 28 01.8 &   \\ 
\hline
\end{tabular}
\end{table}

\subsection{Galactic sources}\label{s_galactic}
The Galactic plane $|b|<1.5\degr$ was excluded from the main AT20G follow-up. However,
as part of our analysis some sources at higher Galactic latitude have been identified as 
Galactic Planetary Nebulae or HII regions. These have been identified in the source 
catalogue (with flags {\it p} and {\it h} respectively) to allow them to be excluded 
from extragalactic studies.

The 65 sources identified as PNe were found by looking for flat spectrum sources that 
were slightly extended, in terms of the 6~km visibility data. A full analysis including 
this data will be published separately. The 6 HII regions were identified through a 
search of the literature.

\section{Completeness and reliability}\label{s_comp}

The overall completeness of the AT20G catalogue is a function of both the completeness of 
the original scanning survey, and the completeness of the follow-up survey.
The methods for estimating completeness for both surveys are discussed in this section. 
In Section~\ref{s_wmap} we compare our catalogue with the \citet{wright09} WMAP catalogue to
assess completeness and reliability. 
\begin{figure*}
\centering
\includegraphics[width=12cm,angle=180]{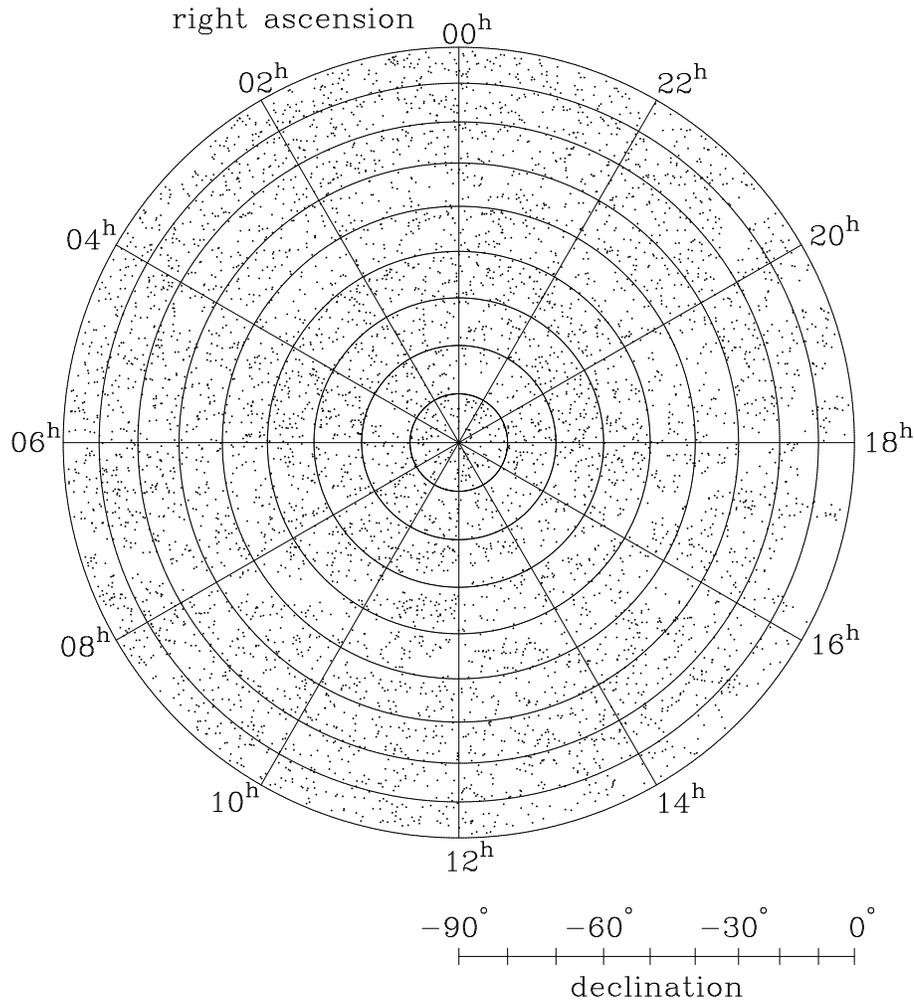}
\caption{A plot in the equal-area Lambert projection showing the distribution of the 5890 sources 
in the AT20G catalogue. Note that the catalogue excludes the Galactic plane ($|b|<1.5\degr$) and 
that some regions north of $\delta=-15\degr$ are incomplete due to bad weather.\label{f_srcs}}
\end{figure*}

\subsection{Scanning survey completeness}
In order to estimate the completeness of the survey catalogue (the catalogue of candidate 
sources that was used as a basis for the follow-up observations) false point sources with known
brightness and positions were inserted into the time ordered uncalibrated scans. 
The false sources were injected by taking the
primary calibrator observation, scaling it to the correct flux and
declination, applying an inverse calibration appropriate to the
time of the observation, and adding this to the raw data. The data was then
processed using the same machinery that created the initial sky maps
and a list of candidate sources including the injected sources were then detected.
Finally, the extracted sources were compared to the list of input sources to measure
the fraction of sources recovered as a function of flux density.
Further details of this process are discussed in Hancock et al. (in preparation).

\changes{This process allowed us to assess the completeness for point sources. The AT20G survey 
is not sensitive to sources with angular sizes much larger than $45\arcsec$ and so our catalogue 
will be incomplete for extremely extended sources.
This brightness sensitivity limit means the AT20G survey can not detect the emission 
from most nearby spiral galaxies, even those with integrated flux densities above 100\,mJy at 
20\,GHz. The nearby spiral NGC\,253 is detected as an extended AT20G source (J004733$-$251717) 
with a 20\,GHz flux density of 608\,mJy, and in this case the observed emission appears to 
arise from a central starburst rather than an AGN \citep[see][]{tingay04,brunthaler09}.
At z=0.0008 (d=3.3\,Mpc), NGC\,253 is the lowest-redshift galaxy detected 
in the AT20G survey.  Although several other spiral galaxies are detected, including NGC\,1068 
(J024240$-$000046; z=0.0038), NGC\,4594 (J123959$-$113721; z=0.0036) and NGC\,4945 
(J130527$-$492804; z=0.0019), the detected radio emission in these galaxies appears to be associated 
mainly with an active nucleus rather than processes related to star formation.}

\subsection{Follow-up survey completeness}
The completeness of the follow-up survey is a function of the completeness of
the scanning survey catalogue, and of the number of objects that were 
scheduled and observed in follow-up mode. 
The number of sources that we were able to observe was limited by the
finite length of the project and the weather conditions during each observing run.
The brightest sources in each region of the sky were observed first so 
as to maximize the number of sources confirmed and thus conserve the 
completeness of the follow-up observations. 
Bad weather and occasional hardware malfunctions meant that even though a source was 
scheduled for observation no good data was obtained. Much of this missing 
data was able to be recouped at a later stage via clean up observations, but 
not all. 

Fig.~\ref{f_srcs} shows the source distribution for the AT20G follow-up
survey source catalogue. \changes{We calculated the follow-up survey completeness ($C_f$)
using
\begin{equation}
C_f = \frac{C_s n_f}{n_s}
\end{equation}
where $C_s$ is the completeness of the scanning survey (estimated using the source injection method
described above); $n_f$ is the number of sources detected in the follow-up survey; and $n_s$ is the 
number of real sources expected from the scanning survey.}
Note that the flux densities determined from the original scanning survey were 
accurate to $\sim20$ per cent. Hence towards the flux density cutoff there is
a reduction in completeness due to some sources being omitted from the
original candidate lists.

For regions south of $\delta = -15\degr$ we estimate the survey completeness
to be 91 per cent above 100~mJy and 79 per cent above 50~mJy. 
The region $-15\degr < \delta < 0\degr$ is notably less complete than the 
rest of the survey. This region was observed last, and was hampered by bad
weather. We had several catch-up runs but this was not enough to fill in 
the missing area within the time limits we had set for near-simultaneous 
follow-up. The main region this affected was between 14~hr and 20~hr in right
ascension.
A full analysis of the completeness statistics, including values for each declination
band with $\sim2$~hr right ascension zones will be presented in 
Massardi et al. (in preparation).

Since every object detected in the follow-up survey passed extensive quality
control criteria, in addition to visual inspection, 
the reliability of the catalogue is essentially 100 per cent.

\section{Comparison with the WMAP 5-year source catalogue}\label{s_wmap}
Foreground point-source removal is an important step in the analysis of CMB 
anisotropy data, since the strongest sources need to be identified and masked 
out of the CMB images.
The WMAP team have therefore produced a catalogue of bright 
point sources in the WMAP sky maps at 23, 33, 41, 61 and 94~GHz \citep{wright09}
so that these can be masked out for CMB analysis.  

Since the AT20G survey overlaps in frequency with the lowest WMAP band (20--25~GHz), 
and the AT20G has significantly better sensitivity and resolution 
than the WMAP images, the AT20G catalogue provides an independent check of the 
completeness and reliability of the WMAP point-source catalogues. 

\citet{massardi09} have recently published a detailed analysis 
which uses data from the AT20G Bright Source Sample \citep{massardi08}
to evaluate several different strategies for foreground source detection in the 
WMAP 5-year maps.  The release of the full AT20G catalogue will allow 
similar studies to push below the 0.5~Jy BSS limit and provide deeper complete 
samples for Planck and other CMB experiments. 

A detailed comparison of the AT20G and WMAP source catalogues is beyond 
the scope of this paper, so we focus here on two questions:  (i) what is 
the completeness and reliability of the WMAP source catalogue? and (ii) 
how consistent are the AT20G and WMAP flux density scales at 20--25~GHz?

\subsection{Matching the AT20G and WMAP source catalogues} 
The WMAP 5-year source catalogue \citep{wright09} contains 390 sources, 
of which 186 are in the southern hemisphere.
\citet{wright09} note that the WMAP catalogue is expected to be complete 
for sources stronger than 2~Jy in regions of the sky away from the Galactic plane, 
but also contains some sources with flux densities as low as 0.5~Jy at 23~GHz 
(WMAP K-band)\footnote{\citet{wright09} note that sources with flux densities 
below 1~Jy are unlikely to be detected by WMAP unless they have `benefited' 
from a positive noise or CMB fluctuation.  This in turn leads to a bias in the 
WMAP catalogue at low flux densities \citep[see, e.g.][]{eddington13,jauncey68}.}. 

In matching the final AT20G catalogue with the WMAP 5-year source 
catalogue, we adopted the same 21.35\,arcmin cutoff radius used by 
\citet{massardi08} for the AT20G Bright Source Sample. 
180 of the 186 southern sources in the WMAP catalogue were matched 
with at least one AT20G source within this 21.35\,arcmin radius.   
Based on our determination of the surface density of bright AT20G sources, we expect 
all these matches to be genuine associations. 

In 16 cases (i.e. 9 per cent of the WMAP source catalogue), two or more AT20G sources 
make a significant ($>10$ per cent) contribution to the total flux density in the 
$0.93^\circ$ WMAP beam at 23~GHz. Some of these correspond to AT20G detections of 
several components (e.g. core and hotspot, or two hotspots) of a single extended 
radio galaxy, while others appear to be unrelated pairs of sources. 

\begin{figure}
\centering
\includegraphics[width=\columnwidth]{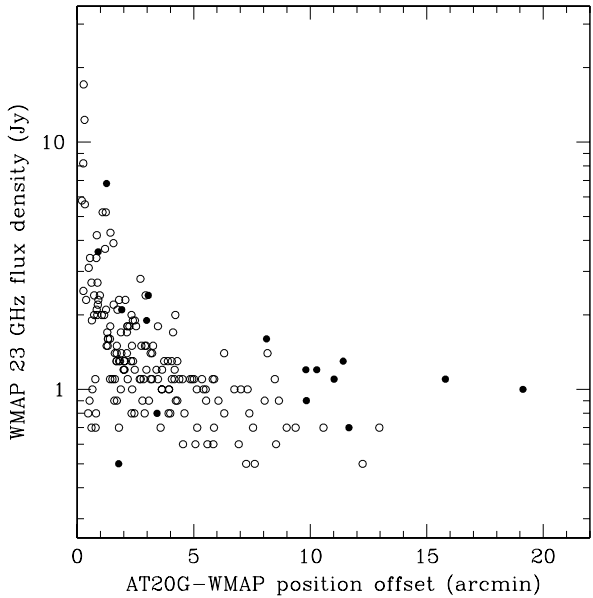}
\caption{Offset between AT20G and WMAP positions for southern sources in 
the 5-year WMAP source catalogue \citep{wright09}.  Open circles 
show WMAP sources that were matched to a single AT20G source, and 
filled circles objects in which more than one AT20G source makes 
a significant ($>10$ per cent) contribution to the flux density in the WMAP 
beam at 23~GHz.
}
\label{wmap5d}
\end{figure}

Fig.~\ref{wmap5d} shows the offsets between the AT20G and WMAP 
positions for sources in common. In general these are consistent with 
the quoted position errors (typically $\sim$4\,arcmin for WMAP 
and less than 1\,arcsec for AT20G).  As expected, the offsets are generally 
smaller for stronger WMAP sources. 

\subsection{Completeness and reliability of the WMAP 5-year source catalogue}
\changes{A simple comparison of surface densities shows that the WMAP and AT20G 
catalogues have similar levels of completeness for strong ($>$1.0\,Jy) 
sources at 20--25~GHz.
The surface density of sources stronger than 1.0\,Jy is 28.5$\pm2.2$\,sr$^{-1}$ 
for AT20G and 26.8$\pm1.7$\,sr$^{-1}$ for the WMAP K-band, where the quoted 
errors are set by the sample size in each case\footnote{The 5-year WMAP catalogue 
covers 78 per cent of the sky (the remaining 22 per cent, at low Galactic latitude, 
is masked out) and lists 262 sources with 23\,GHz flux densities $\geq$1.0\,Jy\,beam$^{-1}$ 
and 123 sources with flux density $\geq$1.5\,Jy\,beam$^{-1}$. The AT20G 
catalogue covers 48 per cent of the sky (declination $<0^\circ$ with the 
region $|b|<1.5^\circ$ masked out) and lists 174 sources with 20\,GHz flux 
densities $\geq$1.0\,Jy\,beam$^{-1}$ and 85 sources with flux density 
$\geq$1.5\,Jy\,beam$^{-1}$.}.
Since the WMAP surface density at 23\,GHz is 94 per cent of the AT20G value, 
and the AT20G catalogue is essentially complete for point sources strong than 1\,Jy, 
we estimate that the WMAP catalogue is roughly 94 per cent complete above 1\,Jy 
at 23\,GHz. Excluding the area masked in the WMAP 
source catalogue, we find only one strong ($>1.5$~Jy) AT20G source which is not 
listed in the WMAP catalogue, AT20G~J142432$-$491349.  

Appendix A lists the eight sources in the WMAP catalogue which have no AT20G match 
stronger than 250\,mJy within 21.35 arcmin of the WMAP position. Three of these 
(the radio galaxy Fornax A, a Galactic HII region and the Galactic planetary nebula 
NGC\,7293) are highly-extended sources 
which are resolved out by the 2\,arcmin ATCA beam, and so represent an 
incompleteness in the AT20G sample. Two sources are found in both surveys 
but with a very large position offset, and three sources catalogued by WMAP are 
not confirmed by AT20G and appear to be spurious. }
In general, however, there is very good agreement between the AT20G and WMAP source 
catalogues.  Since the AT20G catalogue is essentially 100\% reliable, and $\sim$98\% 
of 23\,GHz WMAP sources are matched in AT20G, we conclude that the 23\,GHz WMAP 
catalogue is $\sim$98\% reliable.

\subsection{Comparison of the AT20G and WMAP flux density scales} 
\begin{figure}
\centering
\includegraphics[width=\columnwidth]{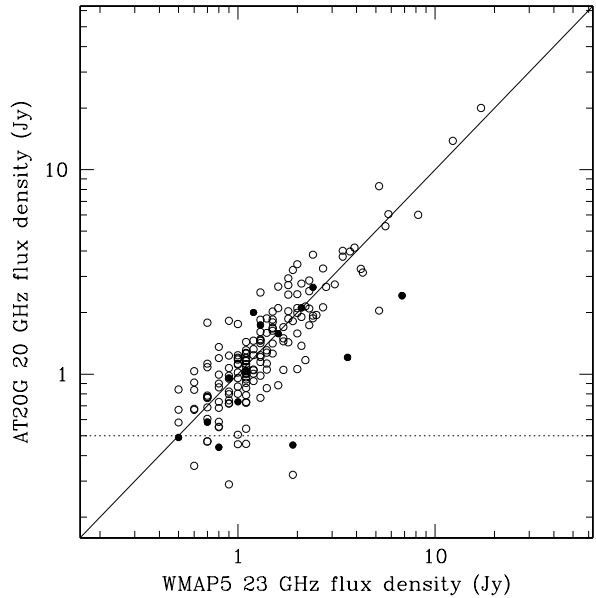}
\caption{Comparison of AT20G (20~GHz) and WMAP (23~GHz) flux densities. 
As in Fig.~\ref{wmap5d}, open circles represent WMAP sources matched 
to a single AT20G source and filled circles objects with two or more 
AT20G sources in the WMAP beam.  The horizontal dashed line shows the 
0.5~Jy limit of the AT20G Bright Source Sample \citep{massardi08}.}
\label{wmap5c}
\end{figure}

Fig.~\ref{wmap5c} compares the catalogued AT20G (20~GHz) and WMAP (23~GHz) 
flux densities for the 180 sources in common. 

To test for consistency of the AT20G and WMAP flux-density scales, 
we compared the flux densities of the 119 sources which were stronger than 
1.0~Jy in the WMAP catalogue (to minimize measurement errors) and had 
only a single AT20G counterpart (to exclude extended sources for which 
the AT20G flux densities may be underestimated).  For these sources, we 
find a mean flux ratio 
\begin{equation}
\left<\frac{{\rm S}_{\rm AT20G}}{{\rm S}_{\rm WMAP}}\right>=1.01\pm0.03 \,\, .
\end{equation}
The rms scatter of the individual fluxes is 0.31.  The main contribution to 
the standard deviation in the flux ratios probably comes from variability,
since the AT20G and WMAP measurements are generally not simultaneous. 
The typical uncertainty in the individual WMAP and AT20G flux density 
measurements is $4-5$ per cent.
Since the formal standard error on the mean flux ratio is 0.028, we conclude that the 
AT20G and WMAP flux density scales are consistent to within 2--3 per cent at 20--25~GHz. 

\section{Analysis}\label{s_prop}

\subsection{Polarisation}\label{s_poln}
\begin{figure}
\centering
\includegraphics[width=\columnwidth]{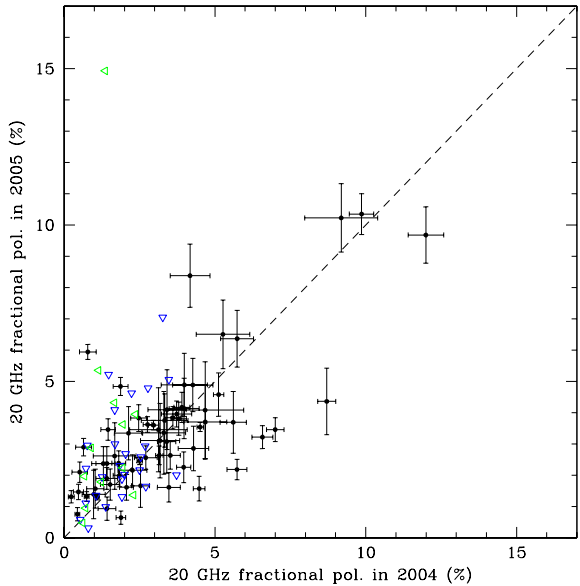}
\caption{Comparison of fractional polarisation at 20~GHz measured in our 2004 October
and 2005 October observing runs. Crosses show measured polarisation
values, and triangles show limits in one of the epochs.}\label{f_poln1}
\end{figure}

To test the robustness of our polarization measurements we used the variability 
sample (see Section~\ref{s_var}), selecting a set of 142 objects which had good 
quality observations in our 2004 October, 2005 October and 2006 April observing
epochs. Fig.~\ref{f_poln1} shows the fractional polarisation at 20~GHz for the
2005 October versus 2004 October observations. Crosses show measured polarisation
values, and triangles show limits in one of the epochs. There is good correlation
(albeit with significant scatter) between each pair of epochs.

\begin{figure}
\centering
\includegraphics[width=\columnwidth]{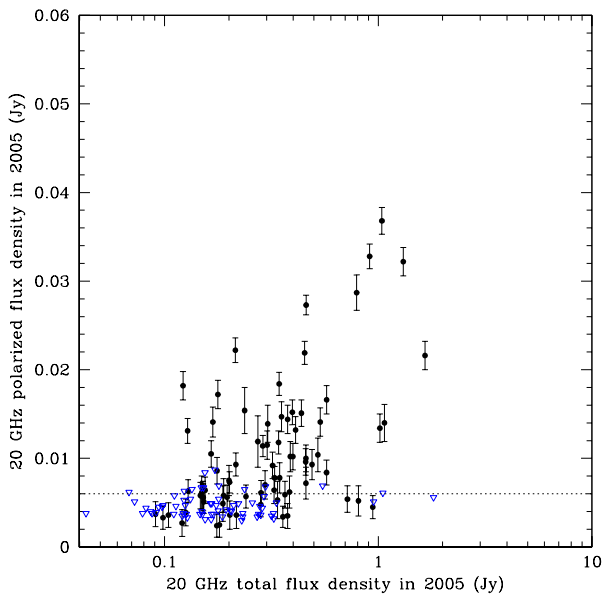}
\caption{Polarised flux density versus total flux density at 20~GHz for the 2005 October 
observing run. Crosses show measured polarisation values, and triangles show 
limits.}\label{f_poln2}
\end{figure}

Fig.~\ref{f_poln2} shows the 20~GHz polarised intensity versus the total flux density
for variability sample sources observed in 2005 October. The dashed horizontal line
shows a 6~mJy limit in polarised flux, a level 3--4 times the typical error in polarised flux.
Most of the limits (triangles) fall below this line, and so it was chosen as a reasonable
level at which the polarised flux density can be reliably determined 
(see Section~\ref{s_polmeas}). Although a small number of sources have a measured polarised 
flux density below 6~mJy, further investigation found that these detections were near the 
$3\sigma$ detection limit, and hence were not highly reliable.

The triangle in the upper left of Fig.~\ref{f_poln1} corresponds to source
J084225--605350, which appeared highly polarized in 2005 (polarized flux $18\pm1$~mJy, 
fractional polarization 5.4 per cent) and 2006 ($16\pm3$~mJy, fractional polarisation
4.0 per cent) but not in 2004 ($<6$~mJy, fractional polarisation $<1.7$ per cent). 
This deserves further investigation, as it raises the possibility that some AT20G sources 
may be genuinely variable in polarized flux.

We found a good correlation between the fractional polarisation measured at 5~GHz 
and 8~GHz, and also a reasonable correlation between the fractional polarisation 
measured at 5~GHz and 20~GHz,
although the large scatter in the latter case suggests there may be some change 
in the polarisation properties of the sources across the wider frequency range.
There was also reasonable consistency in polarised position angles measured at
each of the three frequencies. This will be analysed in more detail in 
Massardi et al. (in preparation).

\subsection{Source counts}
\begin{figure}
\includegraphics[width=\columnwidth]{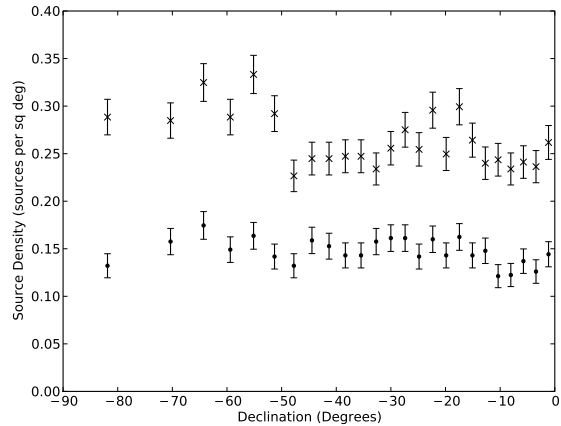}
\caption{Source density in equal area declination bands. Circles and crosses show sources 
above a 100~mJy and 50~mJy cutoff respectively. Errors are Poissonian. At a 50~mJy cutoff we 
have regions of incompleteness (notably 0\degr to $-15\degr$ and $-30\degr$ to $-50\degr$),
as discussed in the text.}\label{f_density}
\end{figure}

Fig.~\ref{f_density} shows the 20~GHz source density with in equal area declination
bands, for flux density cutoffs of 100~mJy and 50~mJy.
It shows that we have relatively homogeneous coverage of the southern sky down to 100~mJy.
At a 50~mJy cutoff we have regions of incompleteness (notably 0\degr to $-15\degr$ and 
$-30\degr$ to $-50\degr$) as discussed in Section~\ref{s_comp}. This is primarily due
to bad weather in the follow-up runs.

\begin{figure}
\includegraphics[width=\columnwidth]{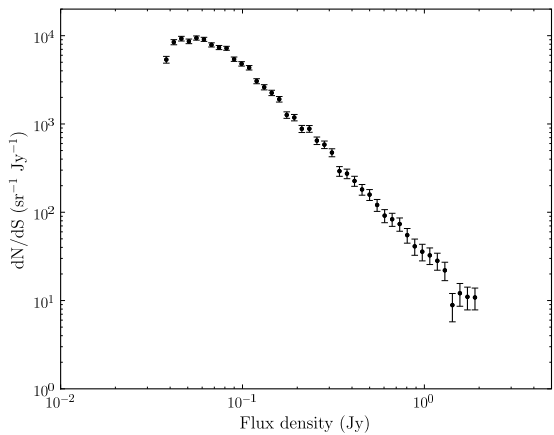}
\caption{$\log(N)-\log(S)$ source counts at 20~GHz, with Poisson errors. These are well 
fit by a power law down to $\sim100$~mJy where the turnover indicates incompleteness.}\label{f_logn}
\end{figure}

Fig.~\ref{f_logn} shows a plot of the AT20G differential source counts ($\log(N)-\log(S)$). These
are well fit by a power law ($f(x) = 31x^{-2.15}$) down to $\sim100$~mJy where the 
flattening of the curve is a sign that the catalogue is becoming incomplete at this level. 
This agrees with our analysis in Section~\ref{s_comp}. Note that the source counts presented here
are not corrected for completeness --- this will be explored further in Massardi et al.,~(in preparation).

\subsection{Spectral index distribution}
\begin{figure}
\includegraphics[width=\columnwidth]{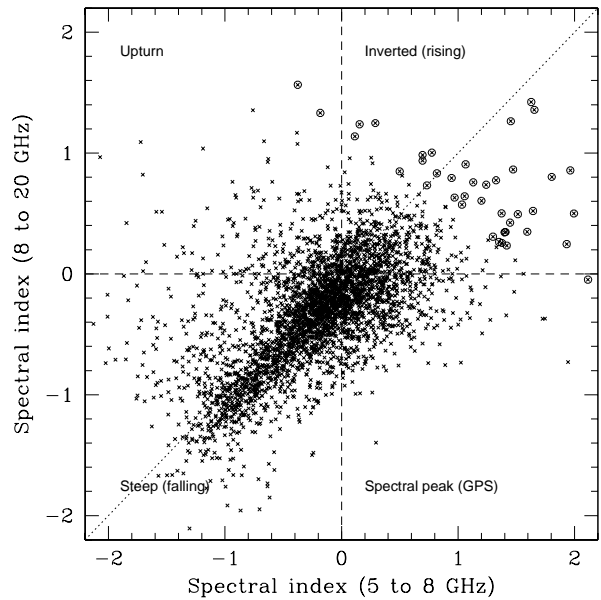}
\caption{Radio two-colour diagram. Crosses show AT20G sources observed at 
all three frequencies. Open circles show the Ultra-Inverted Spectrum (UIS) 
sources with a spectral index of $\alpha(5,20) >+0.7$. These sources are 
discussed in Section~\ref{s_low}.}\label{f_si}
\end{figure}

\changes{
In low-frequency radio surveys a commonly used diagnostic is the spectral 
index:
\begin{equation}
\alpha(\nu_1,\nu_2) =\frac{\log ({S_1}/{S_2})}{\log( {\nu_1}/{\nu_2})}
\end{equation}
where $S_1$ is the flux density measured at frequency $\nu_1$ and likewise for
$S_2$ and $\nu_2$.

The spectral index is only valid over frequency ranges in which the effects of
spectral curvature can be ignored, which is clearly not the case for the AT20G sample.
Hence for our analysis it is more appropriate
to use a radio colour-colour diagram \citep[e.g.][]{kesteven77,sadler06}.
Fig.~\ref{f_si} shows a colour-colour plot for the 3763 AT20G sources with near-simultaneous data at 5, 8 and 20\,GHz. The plot compares the lower frequency
spectral index $\alpha(5,8)$ with the higher frequency spectral index 
$\alpha(8,20)$.

As noted previously by \citet{sadler06} and \citet{massardi08} this plot 
shows a wide range of spectral types at 20~GHz. Four main types can be 
identified:
\begin{enumerate}
\item Sources with steep (falling) spectra, which are shown in the lower 
left quadrant. These represent $\sim 57$ per cent of the sample and 
include a higher fraction of power law spectra.
\item Sources with peaked (GPS) spectra, which rise at lower frequencies 
and fall at higher frequencies. These are shown in the lower right quadrant 
and represent $\sim 21$ per cent of the sample.
\item Sources with inverted (rising) spectra, which are shown in the upper 
right quadrant. These make up $\sim 14$ per cent of sources.
\item Sources with an upturn in their spectra, shown in the upper left 
quadrant. These represent $\sim 8$ per cent of the sample.
\end{enumerate}

The diagonal line indicates sources whose spectrum can be represented by a 
single power law. More sources lie below this line than above it, suggesting 
that most sources steepen with increasing frequency. For the flat spectrum sources 
in particular there is no evidence for power law spectra.  This demonstrates that
high frequency flux densities cannot be reliably estimated from low frequency 
surveys by extrapolating from a single power-law spectrum.

We have also identified a new class of sources with ultra-inverted spectra 
($\alpha(5,20)>+0.7$). These are shown as open circles in Fig.~\ref{f_si} 
and discussed in the next section.}

\subsection{Low--frequency catalogues and Ultra--Inverted Spectrum (UIS) radio sources
}\label{s_low}
To test whether the AT20G survey has found any new radio--source population not 
seen at lower frequency, we cross-matched the AT20G catalogue with the 1.4~GHz 
NVSS \citep{condon98} and 843\,MHz SUMSS \citep{mauch03} and MGPS-2 
\citep{murphy07} surveys, which together cover the whole AT20G survey area. 
The 40\,mJy cutoff of the AT20G catalogue is an order of magnitude brighter 
than the NVSS and SUMSS/MGPS-2 completeness limits (which are typically 
2.5--10\,mJy, depending on declination), so we expect the vast majority of  
AT20G sources to have an NVSS or SUMSS match within 10--15\,arcsec on the sky. 
AT20G sources without a counterpart in the NVSS and SUMSS catalogues must have 
either a sharply--rising radio spectrum (with $\alpha>+0.5$) or highly variable 
radio emission.

Our cross-matching identified 27 AT20G sources (0.4 per cent of the AT20G 
catalogue) with no counterpart in the NVSS and SUMSS/MGPS-2 catalogues, and 
these are listed in Table~\ref{t_no1g}.  For eleven of them, examination of 
the original NVSS and SUMSS survey images shows a weak source at the AT20G 
position --- these objects are detected at the 3$\sigma$ level but fall below 
the NVSS/SUMSS catalogue limit.  For the remaining sixteen objects, 
Table~\ref{t_no1g} lists a 3$\sigma$ upper limit to the low-frequency 
NVSS/SUMSS flux density.

\begin{table*}
\centering
\caption{AT20G sources without a low-frequency counterpart in the SUMSS or NVSS catalogue. The (near-simultaneous) AT20G spectral index between 5 and 20~GHz is also listed.  Optical B(J) identifications and magnitudes are from the Supercosmos catalogue \citep{hambly01},
and infrared K magnitudes from the 2MASS Extended Source Catalog \citep{jarrett00}.
The redshift of J052755$-$471828 has been measured by \citet{drake04} and the redshift of J094258$-$604621 by \citet{radburnsmith06}.
The median 20~GHz flux density of the sources in this table is 72\,mJy. }\label{t_no1g}
\begin{tabular}{lrrrrrrrrll}\hline
          & \multicolumn{5}{c}{Flux density (mJy)} \\
AT20G Name & S20 & S8 & S5 & NVSS & SUMSS & \multicolumn{1}{c}{$\alpha(1,20)$} & \multicolumn{1}{c}{$\alpha(5,20)$} & \multicolumn{1}{c}{$\pm$} & Optical ID & Comment \\
\hline
J023611$-$420337 & 95  & 75  & 36  & ... &   7.3  &  $+$0.81 & $+0.68$ & 0.07 & Faint/blank &  No obvious optical ID \\
J030406$-$450342 & 59  & 49  & 26  & ... & $<3.3$ & $>+0.90$ & $+0.57$ & 0.09 & QSO? & Stellar ID, B(J)=18.0\,mag \\
J032945$-$485420 & 40  & 58  & 47  & ... & $<3.3$ & $>+0.79$ & $-0.11$ & 0.08 & QSO? & Stellar ID, B(J)=19.1\,mag  \\
J034258$-$431813 & 97  & 97  & 39  & ... &   9.4  &  $+0.74$ & $+0.64$ & 0.09 & QSO? & Stellar ID, B(J)=18.4\,mag \\
J044023$-$473218 & 61  & 21  &  9  & ... & $<3.3$ & $>+0.92$ & $+1.34$ & 0.12 & Faint/blank & Possible faint ID, B(J)=22.7 mag \\
J052755$-$471828 & 106 & ... & ... & ... &   9.9  &  $+0.75$ &    ...  &  ..  & Galaxy & 2MASS galaxy K=13.0\,mag, z=0.134 \\ J054417$-$641914 & 123 & 66  & 32  & ... & $<3.3$ & $>+1.14$ & $+0.94$ & 0.08 & Faint/blank & Crowded field near LMC, no obvious ID \\
J070949$-$381152 & 86  & 33  &  6  & $<2.5$ & $<5.1$ & $>+1.33$ & $+1.87$ & 0.15 & Galaxy & 2MASS galaxy with K=12.7\,mag\\
J073040$-$544152 & 70  & ... & ... & ... & $<7.5$ & $>+0.70$ &   ...   &  ..  & Faint/blank  & Low S/N in SUMSS image, no obvious ID \\ J080931$-$472011 & 86  &  50 &  27 & ... &   7.7  & $+0.76$  & $+0.81$ & 0.09 & --- &  Low Galactic latitude, b=$-7.7^\circ$ \\
J094258$-$604621 & 51  & 33  & 23  & ... &   8.7  & $+0.56$  & $+0.59$ & 0.06 & Galaxy & 2MASS galaxy K=12.9\,mag, z=0.096  \\
J095159$-$183703 & 70  & 71  & 38  & $<3.3$ & ... & $>+1.15$ & $+0.43$ & 0.08 & Faint/blank  & No obvious optical ID \\
J104227$-$210556 & 58  & ... &  ...& $<2.0$ & ... & $>+1.27$ &   ...   &  ..  & Faint/blank & No obvious optical ID \\
J111015$-$665531 & 136 & 66  & 21  & ... &   6.0  & $+0.99$  & $+1.31$ & 0.07 & QSO? & Stellar ID B(J)=19.1\,mag \\
J111246$-$203932 & 78  & 30  & 6  & $<2.0$ &  ... & $>+1.38$ & $+1.80$ & 0.17 & Faint/blank &  No obvious optical ID \\
J111605$-$263758 & 40  &  5 &  3  & $<1.2$ &  ... & $>+1.32$ & $+1.82$ & 0.34 & Faint/blank &  No obvious optical ID \\ J114844$-$781933 & 124 & 63  & 22  & ... & $<5.0$ & $>+1.01$ & $+1.21$ & 0.07 & Galaxy & B(J)=18.5\,mag galaxy   \\
J123229$-$840247 & 48  & 86  & 79  & ... &   6.3  & $+0.64$  & $-0.35$ & 0.08 & Faint/blank & No obvious optical ID \\
J140257$-$664031 & 72  & 58  & 26  & ... & $<5.0$ & $>+0.84$ & $+0.71$ & 0.07 & ---  & Low Galactic latitude, b=$-4.8^\circ$ \\
J161845$-$142428 & 84  & ..  & .. & $<2.5$ &  ... & $>+1.32$ &   ...   &  ..  & Faint/blank & No obvious optical ID \\
J171043$-$471820 & 76  & ... & ... & ... &   6.8  & $+0.76$  &   ...   &  ..  & ---  & Low Galactic latitude, b=$-4.4^\circ$ \\
J173039$-$211112 & 54  & ... & ... & $<2.5$ & ... & $>+1.16$ &   ...   &  ..  & Faint/blank  & No obvious optical ID \\
J181225$-$712006 & 43  & 39  & 26  & ... &   5.2  & $+0.67$  & $+0.35$ & 0.09 & Galaxy & 2MASS galaxy K=12.9\,mag \\
J195906$-$755202 & 42  & 30  & 23  & ... & $<3.5$ & $>+0.78$ & $+0.42$ & 0.07 & QSO? & Stellar ID, B(J)=20.6\,mag \\
J200012$-$474951 & 78  & 64  & 28  & ... &   7.4  & $+0.74$  & $+0.72$ & 0.10 & QSO? & Stellar ID, B(J)=20.2\,mag  \\
J205503$-$635207 & 44  & 29  & 12  & ... &   4.6  & $+0.71$  & $+0.91$ & 0.09 & QSO? & Stellar ID, B(J)=20.2\,mag  \\
J220413$-$465424 & 99  & 103 & 30  & ... & $<4.0$ & $>+1.01$ & $+0.84$ & 0.08 & Faint/blank  & No obvious optical ID \\
\hline
\end{tabular}
\end{table*}

Recent results from the 15~GHz 9C survey also confirm a population
of high-frequency sources that fall below the limits of low-frequency 
catalogues. At flux densities of 10--15~mJy, \citet{waldram09} find that 
4.3 per cent of their 15~GHz sources are not listed in NVSS.

All the sources in Table~\ref{t_no1g} have (non-simultaneous) 1--20~GHz 
spectral indices $\alpha(1,20)>+0.7$. An obvious question is whether these 
extreme spectral indices are the result of source variability between the 
NVSS/SUMSS and AT20G observing epochs. This does not generally appear to 
be the case.  The sources in Table~\ref{t_no1g} are also among the most extreme 
AT20G objects in terms of their (simultaneous) 5--20~GHz spectral index, 
with a median value of $\alpha(5,20)=+0.68$ (the median  $\alpha(5,20)$ for 
the AT20G sample as a whole is $-0.22$).  In all but two cases 
(J032945$-$485420 and J123229$-$840247), the data in Table~\ref{t_no1g} are 
consistent with the existence of a population of objects with a rapidly-rising 
power--law radio spectrum between 1 and 20~GHz and little or no variability 
on timescales of up to a decade.

\begin{table*}
\centering
\caption{``Ultra-Inverted Spectrum'' (UIS) AT20G sources with 5--20\,GHz spectral index $\alpha\geq+0.7$.  `He07' 
in the Notes column indicates a source which is also in the CRATES sample of \citet{healey07}. Identifications 
and redshifts are from the NASA Extragalactic Database (NED). Redshift references are: Su04 = \citet{sulentic04};
St91 = \citet{stocke91}; 6dFGS = \citet{jones09}; 2dFGRS = \citet{colless01}. The strongest source in this table, 
J132649$-$525623 (=PMN\,J1326$-$5256) has been observed to show intra-day variability at 6.6\,GHz \citep{mcculloch05}.}
\label{t_uis}
\begin{tabular}{lrrrrrrrrllll}
\hline
          &   \multicolumn{2}{c}{J2000}  & \multicolumn{6}{c}{Flux density (mJy)} \\
AT20G Name & \multicolumn{1}{c}{RA}  & \multicolumn{1}{c}{Dec} & S20 & $\pm$ & S8 & $\pm$ & S5 & $\pm$ & \multicolumn{1}{c}{$\alpha(5,20)$} & \multicolumn{1}{c}{$\pm$} & Notes  \\
\hline
J002616-351249 & 00 26 16.40 & -35 12 49.3 & 1123 & 43 & 357 & 18 & 136 & 7 & +1.48 & 0.06 & PMN\,J0026-3512, He07 \\ 
J011102-474911 & 01 11 02.93 & -47 49 11.3 & 83 & 4 & 64 & 3 & 30 & 2 & +0.71 & 0.08 & z=0.154 galaxy, 2dFGRS  \\ 
J012457-511316 & 01 24 57.38 & -51 13 16.0 & 745 & 37 & 369 & 19 & 229 & 11 &  +0.83 & 0.07 & PKS\,0122-514, He07 \\ 
J024709-281049 & 02 47 09.00 & -28 10 49.7 & 133 & 9 & 108 & 6 & 35 & 2 &  +0.94 & 0.09 & \\ 
J025055-361635 & 02 50 55.42 & -36 16 35.3 & 340 & 16 & 219 & 11 & 84 & 4 &  +0.98 & 0.07 & z=1.536 QSO, Su04 \\ 
J042810-643823 & 04 28 10.87 & -64 38 23.6 & 326 & 15 & 201 & 10 & 110 & 6 &  +0.76 & 0.07 &  PMN\,J0428-6438, He07 \\ 
J043445-421108 & 04 34 45.34 & -42 11 08.0 & 165 & 7 & 58 & 4 & 53 & 3 &  +0.80 & 0.07 &  PMN\,J0434-4211, He07 \\ 
J044023-473218 & 04 40 23.75 & -47 32 18.5 & 61 & 3 & 21 & 2 & 9 & 1 &  +1.34 & 0.12 & \\ 
J050732-510416 & 05 07 32.51 & -51 04 16.3 & 103 & 5 & 31 & 2 & 12 & 1 &  +1.51 & 0.10 & z=0.522 galaxy, St91 \\ 
J051321-212821 & 05 13 21.17 & -21 28 21.4 & 77 & 5 & 33 & 2 & 21 & 2 &  +0.91 & 0.12 & z=0.355 galaxy, 6dFGS \\ 
J054223-514257 & 05 42 23.47 & -51 42 57.4 & 128 & 7 & 69 & 4 & 45 & 2 &  +0.73 & 0.07 & \\ 
J054417-641914 & 05 44 17.81 & -64 19 14.4 & 123 & 6 & 66 & 3 & 32 & 2 &  +0.94 & 0.08 & \\ 
J070949-381152 & 07 09 49.68 & -38 11 52.7 & 86 & 3 & 33 & 3 & 6 & 1 &  +1.87 & 0.15 & K=12.7 galaxy, 2MASS \\ 
J073940-291118 & 07 39 40.11 & -29 11 18.2 & 122 & 8 & 80 & 4 & 25 & 2 &  +1.11 & 0.11 & \\ 
J074109-544746 & 07 41 09.25 & -54 47 46.1 & 86 & 5 & 39 & 2 & 26 & 2 &  +0.84 & 0.10 & K=13.9 galaxy, 2MASS \\ 
J080931-472011 & 08 09 31.97 & -47 20 11.2 & 86 & 4 & 50 & 3 & 27 & 2 & +0.81 & 0.09 & \\ 
J083046-170635 & 08 30 46.57 & -17 06 35.2 & 235 & 15 & 141 & 7 & 70 & 4 &  +0.85 & 0.09 & \\ 
J083529-595311 & 08 35 29.00 & -59 53 11.4 & 549 & 27 & 281 & 14 & 162 & 8 &  +0.86 & 0.07 & PMN\,J0835-5953, He07 \\ 
J095633-404454 & 09 56 33.21 & -40 44 54.8 & 75 & 4 & 39 & 2 & 18 & 1 & +1.00 & 0.08 & \\ 
J101112-221644 & 10 11 12.80 & -22 16 44.5 & 55 & 4 & 29 & 2 & 15 & 2 &  +0.91 & 0.15 & PMN\,J1011-2216   \\ 
J111015-665531 & 11 10 15.79 & -66 55 31.9 & 136 & 6 & 66 & 3 & 21 & 1 & +1.31 & 0.07 & \\ 
J111246-203932 & 11 12 46.81 & -20 39 32.1 & 78 & 5 & 30 & 2 & 6 & 1 &  +1.80 & 0.17 & \\ 
J111605-263758 & 11 16 05.96 & -26 37 58.5 & 40 & 3 & 5 & 1 & 3 & 1 &  +1.82 & 0.34 & \\ 
J114844-781933 & 11 48 44.32 & -78 19 33.6 & 124 & 6 & 63 & 3 & 22 & 1 &  +1.21 & 0.07 & \\ 
J132649-525623 & 13 26 49.23 & -52 56 23.6 & 2061 & 103 & 1350 & 68 & 606 & 30 &  +0.86 & 0.07 & PMN\,J1326-5256 \\ 
J140257-664031 & 14 02 57.40 & -66 40 31.3 & 72 & 4 & 58 & 3 & 26 & 1 &  +0.71 & 0.07 & \\ 
J143608-153609 & 14 36 08.09 & -15 36 09.1 & 113 & 6 & 53 & 3 & 10 & 1 &  +1.70 & 0.11 & \\ 
J144555-303705 & 14 45 55.96 & -30 37 05.5 & 241 & 12 & 105 & 5 & 70 & 5 &  +0.87 & 0.09 & PMN\,J1445-3036, He07 \\ 
J151726-261820 & 15 17 26.60 & -26 18 20.8 & 219 & 14 & 107 & 9 & 80 & 8 &  +0.71 & 0.12 & PMN\,J1517-2618, He07 \\ 
J153030-220811 & 15 30 30.91 & -22 08 11.8 & 78 & 6 & 35 & 3 & 8 & 5 &  +1.60 & 0.74 & \\ 
J153744-295433 & 15 37 44.26 & -29 54 33.6 & 133 & 7 & 93 & 5 & 40 & 3 &  +0.84 & 0.09 & \\ 
J154644-683728 & 15 46 44.52 & -68 37 28.8 & 506 & 25 & 135 & 7 & 168 & 8 &  +0.77 & 0.07 & PMN\,J1546-6837, He07 \\ 
J155205-242521 & 15 52 05.39 & -24 25 21.5 & 139 & 9 & 104 & 6 & 46 & 4 & +0.77 & 0.11 & PMN\,J1552-2425, He07 \\ 
J161434-354329 & 16 14 34.01 & -35 43 29.6 & 243 & 11 & 93 & 6 & 87 & 5 &  +0.72 & 0.07 & \\ 
J171651-470247 & 17 16 51.68 & -47 02 47.1 & 129 & 6 & 45 & 4 & 38 & 3 & +0.86 & 0.09 & \\ 
J172746-754617 & 17 27 46.16 & -75 46 17.9 & 67 & 3 & 50 & 3 & 22 & 1 &  +0.78 & 0.06 & \\ 
J183923-345348 & 18 39 23.56 & -34 53 48.5 & 313 & 16 & 151 & 8 & 64 & 3 &  +1.11 & 0.07 & \\ 
J191816-411131 & 19 18 16.06 & -41 11 31.3 & 357 & 16 & 116 & 6 & 129 & 7 &  +0.71 & 0.07 & PMN\,J1918-4111, He07\\ 
J195949-441611 & 19 59 49.32 & -44 16 11.1 & 172 & 9 & 80 & 4 & 43 & 2 & +0.97 & 0.07 & PMNM\,195617.0-442457\\ 
J200012-474951 & 20 00 12.95 & -47 49 51.5 & 78 & 5 & 64 & 3 & 28 & 2 &  +0.72 & 0.10 & \\ 
J204107-524242 & 20 41 07.70 & -52 42 42.6 & 55 & 3 & 44 & 3 & 20 & 1 & +0.71 & 0.08 &  PMN\,J2041-5242 \\ 
J205503-635207 & 20 55 03.83 & -63 52 07.0 & 44 & 2 & 29 & 2 & 12 & 1 &  +0.91 & 0.09 &  \\ 
J210457-201101 & 21 04 57.07 & -20 11 01.8 & 102 & 6 & 76 & 4 & 30 & 3 & +0.86 & 0.12 & PMN\,J2105-2011\\ 
J220413-465424 & 22 04 13.06 & -46 54 24.7 & 99 & 5 & 103 & 5 & 30 & 2 &  +0.84 & 0.08 & \\ 
J231347-441615 & 23 13 47.91 & -44 16 15.3 & 138 & 7 & 81 & 5 & 46 & 3 &  +0.77 & 0.08 & \\ \hline
\end{tabular}
\end{table*}

By analogy with the Ultra--Steep spectrum (USS) radio sources with $\alpha<-1.3$ 
\citep{blumenthal79,tielens79} we introduce the term ``Ultra-Inverted Spectrum (UIS) radio source'' 
to describe the class of radio sources with a spectral index $\alpha(5,20)>+0.7$.

The AT20G catalogue contains 45 sources with $\alpha(5,20)>+0.7$ (roughly 1.2\% of the AT20G 
sources with 5 and 8~GHz data), and these are listed in Table~\ref{t_uis}.   A few of these 
sources are detected in linear polarization, confirming that the emission mechanism is non-thermal, 
so we assume for now that all the AT20G UIS objects are extragalactic non-thermal
sources. Further investigation is needed to confirm this, but we note that the UIS sources have 
the same distribution in Galactic latitude as the main AT20G sample and show no concentration 
towards the Galactic plane.

\begin{figure}
\begin{center}
\includegraphics[width=8cm]{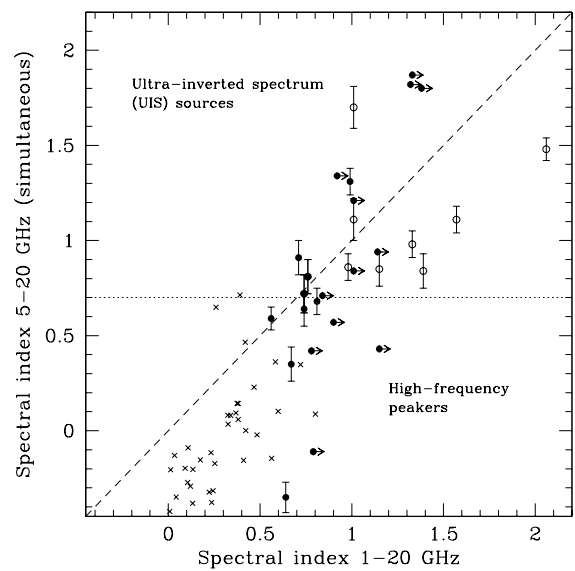}
\caption{Comparison of $1-5$ and $5-20$~GHz spectral indices for the sources in Table~\ref{t_no1g} 
(filled circles) and the `High-Frequency Peaker' sample of \citet{dallacasa00} (crosses).  Open 
circles show AT20G UIS sources (Table~\ref{t_uis}) which have both $\alpha(5,20)>+0.7$ and 
$\alpha(1,20)>+0.7$. \label{f_alpha2}}
\end{center}
\end{figure}

Fig.~\ref{f_alpha2} compares the AT20G sources in Table~\ref{t_no1g} with the sample of High 
Frequency Peakers (HFPs) studied by \citet{dallacasa00}, who measured
simultaneous flux densities at several frequencies between 1.4 and 22.5~GHz for a sample of 55 
sources selected to have inverted spectra with $\alpha(1,5)>+0.5$. Our AT20G UIS sources with 
$\alpha(5,20)>+0.7$ have radio spectra which rise more rapidly with frequency than any of the 
HFP sample.  If the AT20G UIS sources peak at frequencies near or above above 20~GHz, the 
\citet{odea98} relation between spectral peak and source size implies that these are expected 
to be very young, compact radio sources less than a few tens of parsecs in size.

Only a handful of similarly--extreme inverted-spectrum radio sources have previously been 
identified. They include the $z=0.089$ galaxy III Zw 2 \citep{falcke99,brunthaler03} and 
the $z=0.644$ quasar RXJ\,1415$+$3337 \citep{orienti08}.
\citet{falcke99} measured a spectral index of $\alpha=+1.9\pm0.1$ between 5 and 10~GHz for 
III Zw 2, with a spectral peak near 40~GHz, while RXJ\,1415$+$3337 has $\alpha=+0.9$ between 
1 and 22~GHz. Both sources show an evolution of the radio spectrum over timescales of several 
years, with the spectral peak slowly moving to lower frequencies, and III Zw 2 can be modelled 
in terms of a single, adiabatically--expanding homogeneous radio component \citep{orienti08}.

\subsection{Optical Identifications}\label{s_opt}
Optical counterparts of the full AT20G sample were found by cross-matching the radio 
positions with optical positions in the SuperCOSMOS database \citep{hambly01}.
AT20G objects within $10^{\circ}$ of the Galactic plane were excluded from this analysis 
due to the presence of foreground stars and Galactic dust extinction, leaving a total of 
4932 objects used in this analysis. Optical identifications were chosen to be the closest 
optical source to the radio position within a 2.5\,arcsec radius and brighter than a B
magnitude of 22, which is the SuperCOSMOS completeness limit. Previous 
studies have found that 97 per cent of sources selected in this way are likely to be 
genuine associations \citep{sadler06}.

Using this selection method there are 2958 AT20G sources (60 per cent) with optical 
identifications, the majority of which are QSOs. This is much higher than seen in low 
frequency radio surveys which typically have an optical identification rate of 25$-$30 
per cent. This again highlights the significant difference of the AT20G source population 
from other radio selected AGN samples. Full analysis of the optical properties of these 
objects will be published separately (Mahony et al. in preparation)

\section{Conclusions and Future Work}\label{s_conc}
We present a catalogue of 5890 sources from the Australia Telescope 20~GHz
survey, the deepest large-area survey at high radio frequency. For 3766 of
these sources we have near-simultaneous 5 and 8~GHz measurements, and 1559 
sources have a detection in polarized total intensity at one or more of the three frequencies.

The 20~GHz flux densities measured for the strongest AT20G sources are in 
excellent agreement with the WMAP 5-year source catalogue recently published 
by \citet{wright09}, and we find that the WMAP source catalogue is close to complete
(and highly reliable) for sources stronger than 1.5~Jy at 23~GHz. 

We identify a population of Ultra-Inverted Spectrum radio sources with a 
spectral index of $\alpha(5,20)>+0.7$. These are rare sources, comprising 
roughly 1.2 per cent of the AT20G population.

There are several ongoing projects as part of the AT20G. Massardi et al.~(in preparation)
will present a statistical analysis of the AT20G sources, and Hancock et al.~(in preparation)
will present results from the original scanning survey. Mahony et al.~(in preparation) is
carrying out an analysis of the optical identifications of AT20G sources. \citet{sadler08} are
following up subsamples of the AT20G sources at 95~GHz. \citet{murphy09}
is targeting a subset of Galactic sources that were excluded from the main
follow-up survey. Chhetri et al.~(in preparation) has carried out a search for gravitational 
lens candidates and planetary nebulae using data from the 6~km baseline. In a related project, 
Sadler et al.~(in preparation) is conducting a deeper
survey at 20~GHz to explore the high frequency radio population at much lower flux densities.

\section*{Acknowledgments}
We acknowledge the support of the Australian Research Council through 
the award of an ARC Australian Postdoctoral Fellowship (DP0665973) to TM, 
an ARC Australian Professorial Fellowship (DP0451395) to EMS and a 
Federation Fellowship (FF0345330) to RDE. 
Partial financial support for this research has 
been provided to MM and GDZ by the Italian ASI (contracts Planck LFI 
Activity of Phase E2 and I/016/07/0 COFIS) and MUR.

We thank the staff at the Australia Telescope Compact
Array site, Narrabri (NSW), for the valuable support they provide
in running the telescope. The Australia Telescope Compact Array is 
part of the Australia Telescope which is funded by the Commonwealth 
of Australia for operation as a National Facility managed by CSIRO.

This research has made use of the NASA/IPAC Extragalactic Database (NED) which is 
operated by the Jet Propulsion Laboratory, California Institute of Technology, 
under contract with the National Aeronautics and Space Administration.

We thank the referee, Jim Condon, for his useful feedback.

\appendix
\section{WMAP sources missing from the AT20G catalogue} 
Table \ref{wmap1} lists the eight WMAP sources that do not have a 
catalogued bright ($>250$\,mJy) AT20G source within 21.35\,arcmin 
of the WMAP position. These `missing' sources fall into several 
categories which are discussed in the following sections.

\subsection{Bright nearby radio galaxies with large angular size} 
As noted by \citet{massardi08}, the radio galaxy Fornax A 
(WMAP J0322$-$3711) is not detected in the AT20G survey because 
most of its 20~GHz flux density arises from diffuse emission 
associated with the lobes, which are resolved out by the 2\,arcmin 
ATCA beam.  Fornax A is the only bright extragalactic source known to 
be missing from the AT20G catalogue. 

Two other WMAP sources (J0133$-$3627 and J0636$-$2031) correspond to 
peaks in the extended emission of the nearby radio galaxies NGC\,612 
and PKS\,0634$-$20 respectively.  Both these galaxies are detected 
by AT20G, and the large WMAP--AT20G position offsets arise from the 
complex structure of the extended radio emission. \\

\subsection{Extended Galactic sources} 
Two WMAP objects (J0519$-$0539 and J2229$-$2050) are associated with extended 
Galactic sources.  \citet{wright09} note that J0519$-$0539 is a blend 
of two Lynds Bright Nebulae, LBN\,207.65$-$23.11 and LBN\,207.29$-$22.66, 
while J0636$-$2031 is part of the Helix Nebula, NGC\,7293.  
Both objects are significantly larger than the AT20G beam, and appear to 
be mostly resolved out in our survey. \\

\subsection{Extragalactic WMAP sources not found in AT20G} 
There are three remaining WMAP sources (J0632$-$6928, J1150$-$7927 
and J1637$-$7714) for which no obvious counterpart can be found 
in the AT20G catalogue.  All three of these sources are detected 
in each of the five WMAP single-year images (Wright et al.\ 2009), 
and do not appear to vary significantly in flux density over this 
five-year timespan. 

In two cases (WMAP~J0632$-$6928 and J1150$-$7927), there is no catalogued 
AT20G source within the 0.9$^\circ$ WMAP beam at the listed WMAP position.  
Unless these WMAP sources are very diffuse, it seems unlikely  
that they are real. 

J0632-6928 has a WMAP flux density of 0.4\,Jy, making it the weakest 
K-band (23\,GHz) detection in the Wright et al.\ (2009) catalogue, 
but it is detected (as a sub-Jy source) in all 5 WMAP bands.
The closest AT20G source, J063455-694532, is 0.4\,deg away and 
has a flux density of 66\,mJy\,beam$^{-1}$ at 20\,GHz and 154\,mJy\,beam$^{-1}$ at 5\,GHz.  
Its PMN counterpart is PMN J0634-6945 with a flux density of 92$\pm$8\,mJy\,beam$^{-1}$, 
and the SUMSS flux density is 123.1$\pm$3.8\,mJy\,beam$^{-1}$. 
AT20G J063455-694532 seems too faint to be a plausible identification 
for the WMAP source. 

The third source (WMAP J1637$-$7714) is identified by \citet{wright09}
with a nearby 5~GHz source PMN~J1636$-$7713, which is also detected as a weak 
(40$\pm$2\,mJy\,beam$^{-1}$ at 20~GHz) AT20G source.  Once again, however, the catalogued 
AT20G flux density is well below the WMAP detection limit. A stronger AT20G 
source (AT20G~J164416$-$771548, with a 20~GHz flux density of 399$\pm$20\,mJy\,beam$^{-1}$) 
lies 22\,arcmin from the WMAP position, and may be responsible for the WMAP 
detection. 

\begin{table*}
\centering
\caption{The eight southern WMAP sources that are not matched with a 
bright ($>250$\,mJy\,beam$^{-1}$) AT20G source. 
within a matching radius of 21.35\,arcmin.  
The listed WMAP K-band (23~GHz) flux densities are from the 5-year catalogue 
\citep{wright09}.}
\label{wmap1}
\begin{tabular}{lrrcl}
\hline
 WMAP position & WMAP       & $\pm$ & AT20G           & \multicolumn{1}{c}{Notes}  \\ 
   (J2000)     & S$_{23}$ (Jy) & (Jy) & S$_{20}$ (Jy) & \\
\hline
 01 33 26.6 $-$36 27 09  &  0.6 &  0.1 & -- & WMAP source is a hotspot of the radio galaxy NGC\,612, see text   \\
 03 22 25.4 $-$37 11 25  & 18.5 &  3.1 & -- & Radio galaxy Fornax A, see text \\ 
 05 19 21.6 $-$05 39 37  &  2.4 &  0.1 & -- & Blend of Galactic emission nebulae \citep{wright09}, no AT20G counterpart \\
 06 32 21.1 $-$69 28 32  &  0.4 &  0.0 & -- & No AT20G counterpart \\
 06 36 31.8 $-$20 31 38  &  1.1 &  0.0 & 0.13 & Part of the radio galaxy PKS\,0634$-$20, matched with two faint AT20G sources   \\  
 11 50 12.9 $-$79 27 35  &  1.2 &  0.0 & -- & No AT20G counterpart \\ 
 16 37 52.5 $-$77 14 58  &  1.4 &  0.1 & 0.04 & Faint AT20G counterpart, see text   \\
 22 29 47.1 $-$20 50 28  &  0.9 &  0.1 & --   & Part of the Galactic planetary nebula NGC\,7293, no AT20G counterpart \\
\hline
\end{tabular}
\end{table*}

\bibliographystyle{mn2e}
\bibliography{mn-jour,mgps-fix,at20gfsr-fix}

\begin{onecolumn}
\begin{sidewaystable}\label{t_main}
{\bf Table 7.} The first 30 sources in the AT20G source catalogue. \\
\label{t_cat}
\setlength{\tabcolsep}{1mm}
\begin{tabular}{@{}cccrrrrrrcccrrrrrrrrrrrrr}
\hline
IAU designation & $\alpha$ & $\delta$ & 
S$_{20}$ & \(\sigma{_{S_{20}}}\) & S$_8$ & \(\sigma{_{S_8}}\) & S$_5$ & \(\sigma{_{S_5}}\) &
Ep$^a$ & Q$^b$ & F$^c$ & 
P$_{20}$ & \(\sigma{_{P_{20}}}\) & \multicolumn{1}{c}{m} & PA$_{20}$ & 
P$_{8}$ & \(\sigma{_{P_{8}}}\)   & \multicolumn{1}{c}{m} & PA$_{8}$  & 
P$_{5}$ & \(\sigma{_{P_{5}}}\)   & \multicolumn{1}{c}{m} & PA$_{5}$\\

                & (J2000)  & (J2000)  & 
\multicolumn{2}{c}{(mJy)}         & \multicolumn{2}{c}{(mJy)}   &  \multicolumn{2}{c}{(mJy)} &
       &       &      &
\multicolumn{2}{c}{(mJy)} & $(\%)$ & (\degr) & 
\multicolumn{2}{c}{(mJy)} & $(\%)$ & (\degr) & 
\multicolumn{2}{c}{(mJy)} & $(\%)$ & (\degr) \\
\hline
AT20GJ000012$-$853919 & 00:00:12.78 & $-$85:39:19.9 & 98 & 5 & 63 & 4 & 63 & 4 & 222 & g & . & $<$10 & $-$ & $<$9.8 & $-$ & $<$6 & $-$ & $<$9.6 & $-$ & $<$6 & $-$ & $<$9.5 & $-$  \\
AT20GJ000020$-$322101 & 00:00:20.38 & $-$32:21:01.2 & 118 & 6 & 315 & 16 & 515 & 26 & 111 & g & . & $<$6 & $-$ & $<$5.1 & $-$ & $<$6 & $-$ & $<$1.9 & $-$ & $<$6 & $-$ & $<$1.2 & $-$  \\
AT20GJ000105$-$155107 & 00:01:05.42 & $-$15:51:07.2 & 297 & 19 & 295 & 15 & 257 & 13 & 444 & g & . & $<$8 & $-$ & $<$2.7 & $-$ & $<$6 & $-$ & $<$2.0 & $-$ & $<$6 & $-$ & $<$2.3 & $-$  \\
AT20GJ000106$-$174126 & 00:01:06.31 & $-$17:41:26.2 & 73 & 8 & $-$ & $-$ & $-$ & $-$ & 6$\,\cdot$$\,\cdot$ & g & . & $<$6 & $-$ & $<$8.6 & $-$ & $-$ & $-$ & $-$ & $-$ & $-$ & $-$ & $-$ & $-$  \\
AT20GJ000118$-$074626 & 00:01:18.04 & $-$07:46:26.8 & 177 & 9 & $-$ & $-$ & $-$ & $-$ & 6$\,\cdot$$\,\cdot$ & g & . & $<$8 & $-$ & $<$4.4 & $-$ & $-$ & $-$ & $-$ & $-$ & $-$ & $-$ & $-$ & $-$  \\
AT20GJ000124$-$043759 & 00:01:24.50 & $-$04:37:59.6 & 50 & 3 & $-$ & $-$ & $-$ & $-$ & 6$\,\cdot$$\,\cdot$ & g & . & $<$6 & $-$ & $<$12.6 & $-$ & $-$ & $-$ & $-$ & $-$ & $-$ & $-$ & $-$ & $-$  \\
AT20GJ000125$-$065624 & 00:01:25.59 & $-$06:56:24.7 & 77 & 4 & $-$ & $-$ & $-$ & $-$ & 6$\,\cdot$$\,\cdot$ & g & . & $<$8 & $-$ & $<$10.2 & $-$ & $-$ & $-$ & $-$ & $-$ & $-$ & $-$ & $-$ & $-$  \\
AT20GJ000212$-$215309 & 00:02:12.02 & $-$21:53:09.9 & 165 & 11 & $-$ & $-$ & $-$ & $-$ & 4$\,\cdot$$\,\cdot$ & g & . & $<$6 & $-$ & $<$3.6 & $-$ & $-$ & $-$ & $-$ & $-$ & $-$ & $-$ & $-$ & $-$  \\
AT20GJ000221$-$140643 & 00:02:21.71 & $-$14:06:43.9 & 48 & 3 & $-$ & $-$ & $-$ & $-$ & 7$\,\cdot$$\,\cdot$ & g & . & $<$6 & $-$ & $<$12.4 & $-$ & $-$ & $-$ & $-$ & $-$ & $-$ & $-$ & $-$ & $-$  \\
AT20GJ000230$-$033140 & 00:02:30.60 & $-$03:31:40.1 & 53 & 3 & $-$ & $-$ & $-$ & $-$ & 6$\,\cdot$$\,\cdot$ & g & . & $<$9 & $-$ & $<$17.5 & $-$ & $-$ & $-$ & $-$ & $-$ & $-$ & $-$ & $-$ & $-$  \\
AT20GJ000249$-$211419 & 00:02:49.85 & $-$21:14:19.2 & 100 & 7 & $-$ & $-$ & $-$ & $-$ & 4$\,\cdot$$\,\cdot$ & g & . & $<$7 & $-$ & $<$6.9 & $-$ & $-$ & $-$ & $-$ & $-$ & $-$ & $-$ & $-$ & $-$  \\
AT20GJ000252$-$594814 & 00:02:52.93 & $-$59:48:14.0 & 71 & 3 & 64 & 3 & 57 & 3 & 222 & g & . & $<$9 & $-$ & $<$12.2 & $-$ & $<$6 & $-$ & $<$9.4 & $-$ & $<$6 & $-$ & $<$10.5 & $-$  \\
AT20GJ000253$-$562110 & 00:02:53.65 & $-$56:21:10.8 & 94 & 5 & 229 & 12 & 403 & 20 & 222 & g & . & $<$8 & $-$ & $<$8.3 & $-$ & $<$6 & $-$ & $<$2.6 & $-$ & $<$6 & $-$ & $<$1.5 & $-$  \\
AT20GJ000303$-$553007 & 00:03:03.45 & $-$55:30:07.1 & 44 & 3 & 48 & 3 & 52 & 3 & 222 & g & . & $<$7 & $-$ & $<$17.0 & $-$ & $<$6 & $-$ & $<$12.6 & $-$ & $<$6 & $-$ & $<$11.5 & $-$  \\
AT20GJ000311$-$544516 & 00:03:11.04 & $-$54:45:16.8 & 95 & 3 & 313 & 3 & 552 & 9 & 222 & g & e & $<$8 & $-$ & $<$19.9 & $-$ & $<$6 & $-$ & $<$11.8 & $-$ & 15 & 2 & 8.0 & $-$28  \\
AT20GJ000313$-$590547 & 00:03:13.33 & $-$59:05:47.7 & 49 & 3 & 101 & 5 & 151 & 8 & 222 & g & . & $<$8 & $-$ & $<$17.2 & $-$ & $<$6 & $-$ & $<$5.9 & $-$ & $<$6 & $-$ & $<$4.0 & $-$  \\
AT20GJ000316$-$194150 & 00:03:16.06 & $-$19:41:50.7 & 76 & 5 & 162 & 9 & 182 & 9 & 444 & g & . & $<$8 & $-$ & $<$10.6 & $-$ & $<$7 & $-$ & $<$4.4 & $-$ & $<$6 & $-$ & $<$3.3 & $-$  \\
AT20GJ000322$-$172711 & 00:03:22.05 & $-$17:27:11.9 & 386 & 43 & $-$ & $-$ & $-$ & $-$ & 6$\,\cdot$$\,\cdot$ & g & . & 14 & 2 & 3.7 & 18 & $-$ & $-$ & $-$ & $-$ & $-$ & $-$ & $-$ & $-$  \\
AT20GJ000327$-$154705 & 00:03:27.35 & $-$15:47:05.4 & 129 & 8 & $-$ & $-$ & $-$ & $-$ & 4$\,\cdot$$\,\cdot$ & g & . & $<$10 & $-$ & $<$7.5 & $-$ & $-$ & $-$ & $-$ & $-$ & $-$ & $-$ & $-$ & $-$  \\
AT20GJ000404$-$114858 & 00:04:04.88 & $-$11:48:58.0 & 680 & 32 & $-$ & $-$ & $-$ & $-$ & 6$\,\cdot$$\,\cdot$ & g & . & $<$7 & $-$ & $<$1.0 & $-$ & $-$ & $-$ & $-$ & $-$ & $-$ & $-$ & $-$ & $-$  \\
AT20GJ000407$-$434510 & 00:04:07.24 & $-$43:45:10.0 & 199 & 10 & 211 & 11 & 244 & 12 & 111 & g & . & $<$8 & $-$ & $<$4.1 & $-$ & $<$6 & $-$ & $<$2.8 & $-$ & $<$6 & $-$ & $<$2.5 & $-$  \\
AT20GJ000413$-$525458 & 00:04:13.97 & $-$52:54:58.7 & 65 & 3 & 98 & 4 & 192 & 6 & 222 & g & e & $<$7 & $-$ & $<$15.7 & $-$ & $<$6 & $-$ & $<$14.2 & $-$ & $<$6 & $-$ & $<$8.4 & $-$  \\
AT20GJ000435$-$473619 & 00:04:35.65 & $-$47:36:19.0 & 868 & 36 & 970 & 49 & 900 & 45 & 111 & g & . & 15 & 3 & 1.7 & $-$51 & 30 & 2 & 3.1 & $-$45 & 25 & 1 & 2.8 & $-$43  \\
AT20GJ000505$-$344549 & 00:05:05.94 & $-$34:45:49.6 & 131 & 6 & 142 & 7 & 134 & 7 & 111 & g & . & $<$6 & $-$ & $<$4.6 & $-$ & $<$6 & $-$ & $<$4.2 & $-$ & $<$6 & $-$ & $<$4.5 & $-$  \\
AT20GJ000507$-$013244 & 00:05:07.03 & $-$01:32:44.6 & 81 & 4 & $-$ & $-$ & $-$ & $-$ & 6$\,\cdot$$\,\cdot$ & g & . & $<$6 & $-$ & $<$7.8 & $-$ & $-$ & $-$ & $-$ & $-$ & $-$ & $-$ & $-$ & $-$  \\
AT20GJ000518$-$164804 & 00:05:18.01 & $-$16:48:04.9 & 142 & 9 & $-$ & $-$ & $-$ & $-$ & 4$\,\cdot$$\,\cdot$ & g & . & $<$6 & $-$ & $<$4.2 & $-$ & $-$ & $-$ & $-$ & $-$ & $-$ & $-$ & $-$ & $-$  \\
AT20GJ000558$-$562828 & 00:05:58.32 & $-$56:28:28.9 & 151 & 5 & 376 & 5 & 677 & 11 & 222 & g & e & $<$7 & $-$ & $<$7.2 & $-$ & $<$6 & $-$ & $<$8.5 & $-$ & $<$6 & $-$ & $<$2.7 & $-$  \\
AT20GJ000600$-$313215 & 00:06:00.47 & $-$31:32:15.0 & 63 & 4 & 53 & 3 & 52 & 3 & 111 & g & . & $<$9 & $-$ & $<$13.8 & $-$ & $<$6 & $-$ & $<$11.3 & $-$ & $<$6 & $-$ & $<$11.6 & $-$  \\
AT20GJ000601$-$295549 & 00:06:01.14 & $-$29:55:49.6 & 97 & 6 & 187 & 10 & 228 & 10 & 444 & g & . & $<$6 & $-$ & $<$6.2 & $-$ & $<$7 & $-$ & $<$3.9 & $-$ & $<$6 & $-$ & $<$2.6 & $-$  \\
AT20GJ000601$-$423439 & 00:06:01.95 & $-$42:34:39.8 & 110 & 5 & 259 & 14 & 532 & 27 & 111 & g & . & 11 & 3 & 9.8 & 32 & 14 & 1 & 5.5 & 22 & $<$6 & $-$ & $<$1.1 & $-$  \\
\hline
\end{tabular}

\flushleft
NOTES:\\
\(^{\rm{a}}\) Epoch of follow-up observations for 20 GHz, 8 GHz and 5 GHz fluxes respectively. The epochs
are listed in Table 3.\\
\(^{\rm{b}}\) Quality flag: {\bf g} (good) or {\bf p} (poor).
\(^{\rm{c}}\) Other flags: see Section \ref{s_cat} for description.

\end{sidewaystable}

\end{onecolumn}


\label{lastpage}
\end{document}